\documentclass[final,5p,times,twocolumn]{elsarticle}

\usepackage{amssymb}
\usepackage{amsthm}
\usepackage{amsmath,bm}
\usepackage[hidelinks]{hyperref}

\usepackage{graphicx}
\usepackage{geometry}
\usepackage{floatrow}
\usepackage{multirow}
\usepackage{lipsum}

\usepackage{caption}
\usepackage{subcaption}
\usepackage{wrapfig}
\usepackage{comment}
\usepackage{dblfloatfix}
\usepackage{float}
\biboptions{sort&compress}
\usepackage{hyperref}

\makeatletter
\newcommand{\labeltext}[2]{%
  \@bsphack
  \MakeLinkTarget*{#1}%
  \def\@currentlabel{#1}{\label{#2}}%
  \@esphack
}
\makeatother

\journal{Applications in Energy and Combustion Science}

\begin{document}

\begin{frontmatter}

\title{Numerical Simulation of Reacting and Non-Reacting Liquid Jets in Supersonic Crossflow}

\author{Michael Ullman$^*$}
\author{Shivank Sharma}
\author{Venkat Raman}
\address{Department of Aerospace Engineering, University of Michigan, Ann Arbor, MI, USA}

\begin{abstract}
    Canonical jet in supersonic crossflow (JISC) studies have been widely used to study fundamental physics relevant to a variety of applications. While most JISC works have considered injection of gases, liquid injection is also of practical interest and introduces additional multi-scale physics, such as atomization and evaporation, that complicate the flow dynamics. To facilitate further understanding of these complex phenomena, this work presents multiphase simulations of reacting and non-reacting JISC configurations with freestream Mach numbers of roughly 4.5. Adaptive mesh refinement is implemented with a volume of fluid scheme to capture the liquid breakup and turbulent mixing at high resolution. The results compare the effects of the jet momentum ratio and freestream temperature on the jet penetration, mixing, and combustion dynamics. For similar jet momentum ratios, the jet penetration and mixing characteristics are similar for the reacting and non-reacting cases. Mixing analyses reveal that vorticity and turbulent kinetic energy (TKE) intensities peak in the jet shear layers, where vortex stretching is the dominant turbulence-generation mechanism for all cases. Cases with lower freestream temperatures yield negligible heat release, while cases with elevated freestream temperatures exhibit chemical reactions primarily along the leading bow shock and within the boundary layer in the jet wake. The evaporative cooling quenches the chemical reactions in the primary atomization zone at the injection height, such that the flow rates of several product species plateau after $x/d=20$. Substantial concentrations of final product species are only observed along the bow shock—due to locally elevated temperature and pressure—and in the boundary layer far downstream—where lower flow velocities counteract the effects of prolonged ignition delays. This combination of factors leads to low combustion efficiency at the domain exit.
\end{abstract}


\end{frontmatter}


\section{Introduction\label{sec:introduction}}

Over the past century, studies of canonical jet in crossflow configurations have been extensively utilized to investigate fundamental physics relevant to propulsion and power generation applications \cite{KARAGOZIAN2010531, Mahesh2013}. More recently, jet in supersonic crossflow (JISC) setups have received increased attention due to their relevance for high-speed propulsion \cite{liu2020_review, gyusub_frontiers, urzay2018_arfm}. These flowpaths involve complex interactions between a number of physical phenomena, including shock waves, boundary layers, turbulence, and finite-rate chemical kinetics—all of which are dependent on the flow conditions and geometry. As such, simplified JISC configurations have allowed these competing effects to be examined in greater detail.

These studies are primarily aimed at elucidating how injected fuel is able to mix and react with the freestream over short residence times. While low-speed jet in crossflow configurations exhibit several similar features, such as horseshoe, shear-layer, and counter-rotating vortex pairs \cite{KARAGOZIAN2010531, Mahesh2013}, other features, such as bow and separation shocks upstream of the injection point, are unique to supersonic flows. In the past, JISCs with gaseous injection have been studied more extensively both experimentally \cite{gamba2015_jisc, gruber1995_jisc, rothstein1992study, mcclinton1974effect} and computationally \cite{sharma2024_scitech, sharma2024_proci, liu2019characteristics, 2013_Watanabe}. However, several have also considered liquid injection \cite{lin2017_pdpa, sharma2025_aiaa, yates1972liquid, perurena2009experimental}, which introduces additional multi-scale phenomena such as fuel atomization and evaporation. These factors in liquid JISCs lead to increased combustion timescales relative to their gaseous counterparts \cite{bhatia2023}. The underlying processes facilitating fuel atomization and combustion in these scenarios are not well understood.

\begin{figure}[!h]
    \centering
    \includegraphics[width=\textwidth]{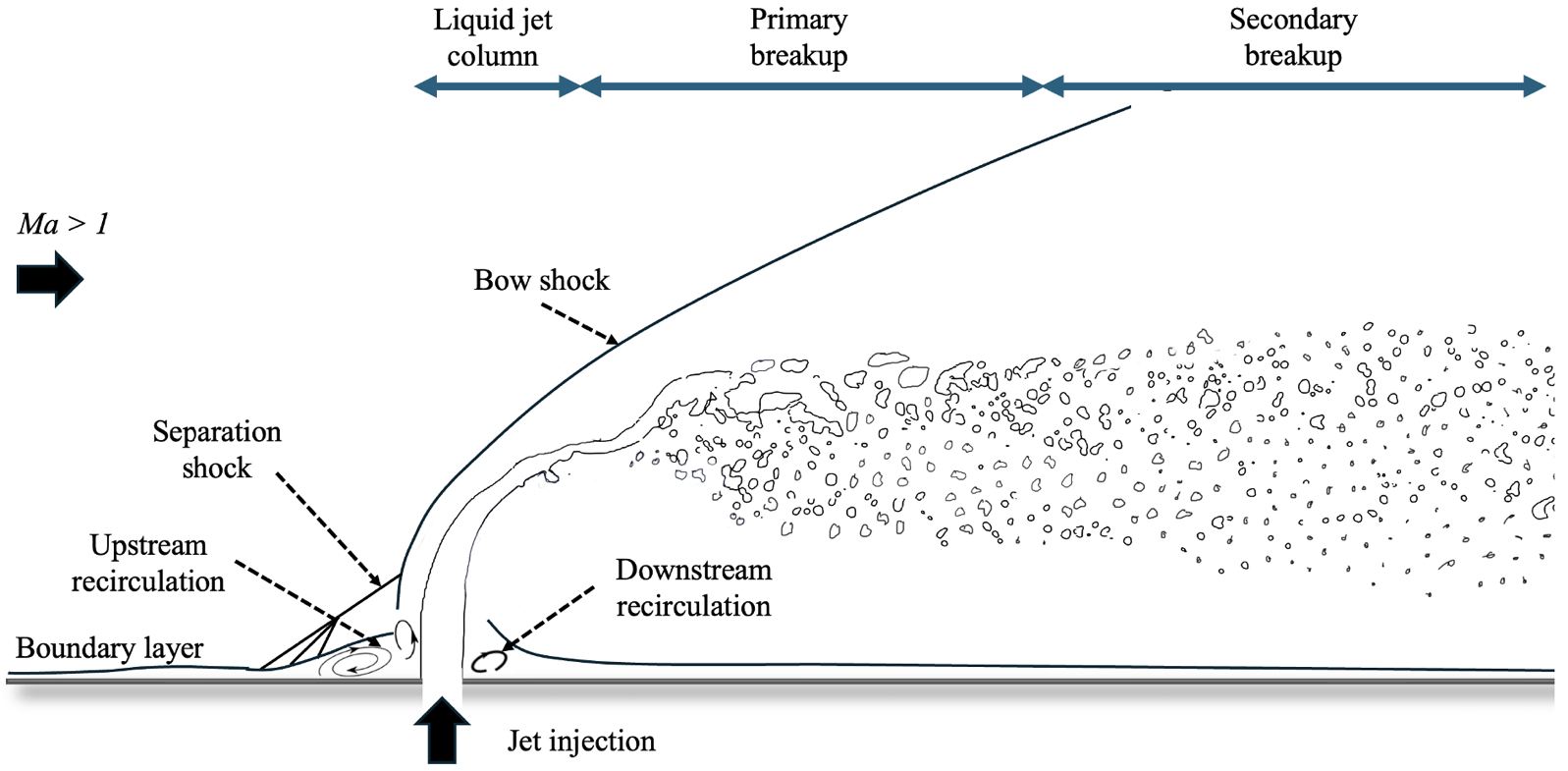}
    \caption{Schematic of a liquid jet injected into a supersonic crossflow \cite{sharma2025_aiaa}.}
    \label{fig:jisc_schematic}
\end{figure}

Figure \ref{fig:jisc_schematic} shows a schematic of a typical liquid JISC configuration. The column of injected liquid obstructs the supersonic freestream, creating a bow shock on the windward side of the jet. This creates an adverse pressure gradient, which facilitates boundary layer separation upstream of the injector. The compression waves created by this separation coalesce into a separation shock, which impinges upon the bow shock from the injector. This can lead to entrainment of fuel into the recirculation zone upstream of the injector, which can then act as a flameholder \cite{gruber1996bow, yakar1998experimental, perurena2009experimental}. Downstream, the outer core of the liquid jet initially breaks up into larger primary droplets, which themselves are further broken up into smaller secondary droplets by aerodynamic forces. These secondary droplets are then able to evaporate and react with the crossflow \cite{ghods2013}. Kelvin-Helmholtz (KH) and Rayleigh-Taylor (RT) instabilities along the jet surface can facilitate the breakup of the liquid column \cite{sallam2004breakup, schetz1980wave, less1985transient} as well as the downstream atomization of the fuel droplets \cite{schetz1980wave, sallam2017digital}. In the jet wake, additional KH and RT instabilities generate shear layer vortices and counter-rotating vortex pairs, which enhance mixing with the crossflow \cite{Yuan1999, Fric1996, gruber1995_jisc, Yakar2006}.

Existing liquid JISC studies have primarily investigated jet penetration heights for different flow conditions and analyzed the various liquid atomization processes. Several experimental techniques have been employed for these purposes, including Schlieren and shadowgraphy \cite{kolpin1968study, boiko2019liquid}, holography \cite{sallam2004breakup, johnson2023digital, cao2021experimental, sallam2017digital, prakash2018_ljisc}, particle image velocimetry (PIV) \cite{boiko2019liquid}, phase Doppler particle analysis (PDPA) \cite{lin2017_pdpa, lin2004_pdpa}, and x-ray diagnostics \cite{lin2020exploration}. While these studies have provided critical insight into the relevant fundamental physics, experimental diagnostics often provide limited access to the internal flow structures and small-scale processes driving the global system behavior. To fill this gap, high-fidelity numerical simulations can be employed to further examine the underlying phenomena.

Current numerical techniques for simulating these problems can be divided into three types. The first is Lagrangian particle tracking (LPT), wherein discrete liquid particles in a Lagrangian reference frame interact with the surrounding gaseous flow, which is simulated in the Eulerian reference frame. This approach is commonly used to study spray combustion and full-scale propulsion devices \cite{vujanovic2016_spray, prakash2024_liqRDE, wang2022_eullag}, but its predictive capabilities are often limited by its need for heat transfer, evaporation, and breakup models \cite{sirignano1983_pecs, musick2023_dropdet}. The second type of numerical approach encompasses interface-tracking techniques, such as level set or ghost-fluid methods, where liquid-gas interfaces are represented as discontinuities within the flow \cite{sharfuddin2021level, pan2018_levelset, liu2011_ghostfluid}. These techniques have the advantage of well-defined phasic interfaces, but can introduce spurious oscillations and conservation issues near interfaces, making them less stable for high-speed multiphase flows \cite{dorschner2020_jfm, fuster2019review, saurel2018review}. The final type of numerical approach is interface-capturing methods, such as volume of fluid (VOF), where phasic interfaces are diffused over a few computational cells in the multiphase Eulerian solution \cite{kuhn2022experimentally, hess2023numerical, bielawski2024analysis, bielawski_thesis, sharma2025_aiaa}. These methods provide increased stability relative to interface-tracking schemes, while also yielding more detailed representations of phasic interfaces, droplet breakup, and evaporation. However, interface-capturing methods generally incur greater computational expense, as comparatively fine computational grids are required to resolve small ligaments and droplets with reasonable accuracy.

To date, computational studies of reacting JISC have primarily considered gaseous jets—typically hydrogen \cite{sharma2024_proci, liu2019characteristics, 2013_Watanabe} or ethylene \cite{kim2012_reactjisc, sharma2024_scitech}. Simulations of liquid JISC have been predominantly conducted with water \cite{kuhn2022experimentally, zhou2023_waterjisc, zhao2022_waterjisc}, as have experiments \cite{kolpin1968_waterjisc, perurena2009experimental, zhao2021_waterjisc, lin2004_pdpa, ghenai2009, yates1972liquid}. As such, existing literature on reacting liquid JISC is sparse, with only a handful of experiments \cite{vinogradov2007review} and simulations \cite{sharma2025_aiaa} having been conducted. To address this, the current study presents VOF simulations of reacting and non-reacting liquid JISC with a freestream Mach number of roughly 4.5. Adaptive mesh refinement is used to efficiently capture the near-injector region, jet wake mixing, and liquid atomization at high spatiotemporal resolution. The results detail the jet penetration, mixing, and combustion characteristics driving the global behaviors of each configuration.


\section{Numerical methods}

The solver is a VOF extension of an in-house compressible reacting flow solver \cite{numericsHMM}, which employs block-structured adaptive mesh refinement (AMR) through the AMReX framework \cite{AMReX_JOSS}. This allows spatiotemporally-evolving flow features to be captured via user-defined refinement criteria. This solver, as well as its gas-phase predecessors \cite{bielawski_2022_umreactingflow, numericsHMM}, have been extensively utilized to study high-speed reacting flows \cite{sharma2025_aiaa, bielawski_thesis, sharma2024_proci, abisleiman2025_odw_cnf, rauch2024_scitech, ullman2024_cnf_strat, ullman2025_m7ai, prakash_pci_rde, ullman_drone_CF}.

\begin{figure*}[!hb]
    \centering
    \includegraphics[width=0.8\textwidth]{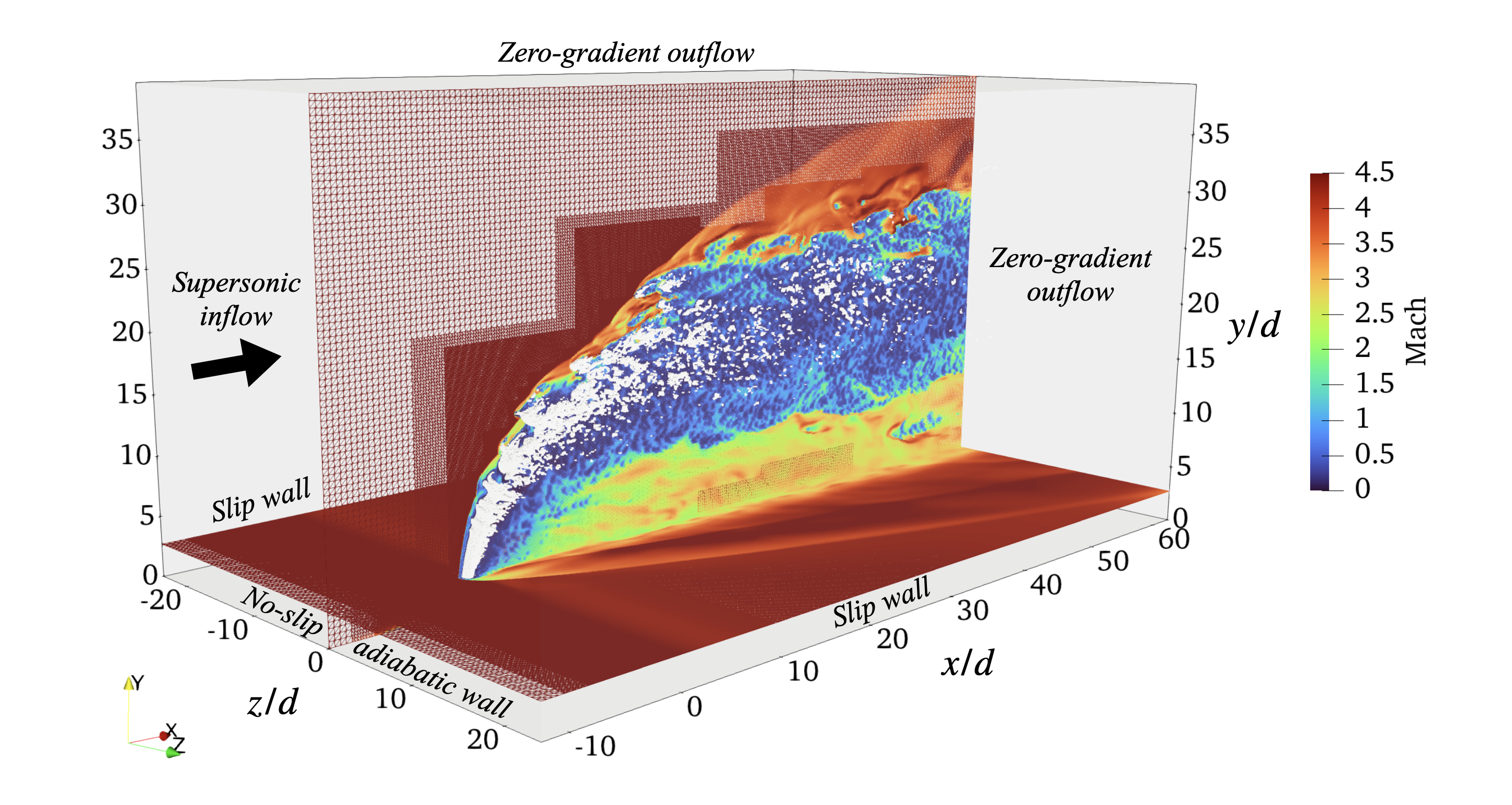}
    \caption{Schematic of boundary conditions for each case, along with liquid volume fraction isocontour (white) and slices of Mach number contours in case \ref{case4_M8}. AMR refinement can be seen from the density of the wireframe in the Mach contours.}
    \label{fig:case_schematic}
\end{figure*}

The multiphase system of equations follows those developed by Saurel et al.\ \cite{saurel2009simple} and extended by Bielawski et al.\ \cite{bielawski2024analysis, bielawski_thesis} to account for phase change and viscous effects. In the VOF methodology, $\alpha_l$ and $\alpha_g$ denote the volume fractions of the immiscible liquid and gas phases within a given computational cell. These phases are respectively governed by the ideal gas and stiffened gas equations of state. The governing equations for liquid volume fraction and conservation of mass, momentum, energy, and chemical species are:
\begin{equation}
    \frac{\partial \alpha_l}{\partial t} + u_j \frac{\partial \alpha_l}{\partial x_j} = \mathcal{R}_{\alpha_l} (p_l - p_g, T_l-T_g, g_l-g_v)
    \label{eqn:vol_frac}
\end{equation}
\begin{equation}
    \frac{\partial \alpha_l \rho_l}{\partial t} + \frac{\partial \alpha_l \rho_l u_j}{\partial x_j} = \mathcal{R}_{\rho_l \alpha_l}(p_l-p_g, T_l-T_g, g_l-g_v)
    \label{eqn:alpha_l}
\end{equation}
\begin{equation}
    \frac{\partial \alpha_g \rho_g}{\partial t} + \frac{\partial \alpha_g \rho_g u_j}{\partial x_j} = \mathcal{R}_{\rho_g \alpha_g}(p_l-p_g, T_l-T_g, g_l-g_v)
    \label{eqn:alpha_g}
\end{equation}
\begin{equation}
    \frac{\partial \rho u_i}{\partial t} + \frac{\partial \rho u_i u_j}{\partial x_j} + \frac{\partial \alpha_l p_l}{\partial x_i} + \frac{\partial \alpha_g p_g}{\partial x_i} - \frac{\partial \tau_{ij}}{\partial x_j} = 0
    \label{eqn:momentum}
\end{equation}
\begin{multline}    
    \frac{\partial \rho E}{\partial t} + \frac{\partial \rho E u_j}{\partial x_j} + \frac{\partial p u_j}{\partial x_j} - \frac{\partial \tau_{ij} u_i}{\partial x_j} - \frac{\partial}{\partial x_j} \left( \lambda \frac{\partial T}{\partial x_j} 
    \right) \\
    = \mathcal{R}_{\rho_g E} (p_l-p_g, T_l-T_g, g_l-g_v)
    \label{eqn:energy}
\end{multline}
\begin{multline}
    \frac{\partial \rho_g \alpha_g Y_k}{\partial t} + \frac{\partial u_j \rho_g \alpha_g Y_k}{\partial x_j} - \frac{\partial}{\partial x_j} \left( \rho_g D \frac{\partial Y_k}{\partial x_j} \right) - \alpha_g \dot{\Omega}_k \\
    = \mathcal{R}_{\rho_g \alpha_g Y_k} (p_l-p_g, T_l-T_g, g_l-g_v), \quad k=1, \dots, N_s
    \label{eqn:species}
\end{multline}
where
\begin{equation}
    \tau_{ij} = \mu \left( \frac{\partial u_i}{\partial x_j} + \frac{\partial u_j}{\partial x_i} \right) - \frac{2}{3} \mu \frac{\partial u_k}{\partial x_k} \delta_{ij}
    \label{eqn:tau_ij}
\end{equation}
\begin{equation}
    e_l = \frac{p+\gamma_l \Pi_{l,\infty}}{(\gamma_l -1)\rho_l} + e_{0,l}
    \label{eqn:e_liq_eos}
\end{equation}
\begin{equation}
    e_g = \int_{T_0}^T C_pdT - \frac{p_g}{\rho_g}+\sum_{k=1}^{N_s}\Delta h^0_{f,k}Y_k
    \label{eqn:e_gas_eos}
\end{equation}
\begin{equation}
    \rho E = \rho_l \alpha_l e_l + \rho_g \alpha_g e_g + \frac{1}{2} \rho u_i u_i
    \label{eqn:rhoE}
\end{equation}

In these equations, the mixture density ($\rho$), viscosity ($\mu$), and thermal conductivity ($\lambda$) are computed using volumetric mixture rules with components from each phase. The chemical source term for the $k$-th species, $\dot{\Omega}_k$, is only integrated where the gas phase is dominant—i.e., where $\alpha_l < \alpha_{min}$. In this work, the threshold for identifying single-phase flow, $\alpha_{min}$, was set to $10^{-10}$. The relaxation operators $\mathcal{R}$ in Eqs.\ \ref{eqn:vol_frac}-\ref{eqn:species} are each functions of the pressure ($p$), temperature ($T$), and Gibbs free energy ($g$) of each of the phases. Within each simulation timestep, a relaxation routine is performed such that the relaxation operators tend toward zero, and the phases are taken to be in mechanical, thermal, and chemical equilibrium. This allows for evaporation of the liquid phase or condensation of the vapor species (treated simply as one of the gaseous species) within a given timestep. The reader is referred to Ref.\ \cite{bielawski_thesis} for further details on the vapor-liquid equilibrium algorithm.

A second-order accurate finite-volume discretization is used to solve the governing equations. The convective fluxes are computed using the Harten-Lax-van Leer-Contact (HLLC) scheme, while the diffusive fluxes are computed using central differences. Temporal integration is performed using an explicit two-stage Runge-Kutta scheme \cite{shu_rk}. The nonlinear reconstruction method $\rho$-THINC \cite{garrick2017_rhothinc} is used to prevent excess dissipation of phasic interfaces. The gaseous thermodynamic state and chemical source terms are computed by Cantera \cite{goodwin2020cantera} with a 43 species, 336 reaction mechanism developed using the Foundational Fuel Chemistry Model \cite{ffcm2, dodecane_mechanism}.

Table \ref{table:sg_parameters} provides the stiffened gas parameters for the liquid phase, which were calibrated from NIST data over the expected pressure and temperature ranges \cite{nist_fluid_props}. This calibration was performed such that the liquid density and speed of sound from stiffened gas equation of state matched the NIST data at the freestream pressure (20 kPa) and plenum temperature (300 K). When computing the liquid and mixture states, the parameters in Table \ref{table:sg_parameters} are held constant. Similarly, the liquid viscosity and thermal conductivity are held constant with values taken from NIST fluid data at 300 K \cite{nist_fluid_props}.

\begin{table}[h]
\centering
\caption{Stiffened gas parameters.}
\begin{tabular}{|c|c|c|c|}
\hline
\textbf{$\boldsymbol{c_{v,l}}$} & \textbf{$\boldsymbol{\gamma_l}$} & \textbf{$\boldsymbol{\Pi_{l,\infty}}$} & \textbf{$\boldsymbol{e_{0,l}}$} \\ \hline \hline
904.19 & 2.98981 & 4.014 $\times 10^8$ & -2.998 $\times 10^6$ \\ \hline
\end{tabular}
\label{table:sg_parameters}
\end{table}

\begin{table*}[!h]
\centering
\caption{Case configurations.}
\begin{tabular}{|c|c|c|c|c|c|c|c|c|}
\hline
\textbf{Case} & \textbf{$\boldsymbol{T_{\infty}}$ {[}K{]}} & \textbf{$\boldsymbol{p_{\infty}}$ {[}kPa{]}} & \textbf{$\boldsymbol{M_{\infty}}$} & \textbf{$\boldsymbol{p_{plenum}}$ {[}MPa{]}} & \textbf{$\boldsymbol{T_{plenum}}$ {[}K{]}} & \textbf{$\boldsymbol{J}$} & \textbf{Reactions} & \textbf{No. Cells}\\ \hline \hline
1\labeltext{1}{case2} & 256 & 21.7 & 4.51 & 3.87 & 300 & 11.73 & Off & $1.96 \times 10^7$ \\ \hline
2\labeltext{2}{case4} & 246 & 21.2 & 4.53 & 7.54 & 300 & 23.95 & Off & $2.73 \times 10^7$ \\ \hline
3\labeltext{3}{case2_M8} & 702 & 21.7 & 4.51 & 3.87 & 300 & 12.22 & On & $2.30 \times 10^7$ \\ \hline
4\labeltext{4}{case4_M8} & 669 & 21.2 & 4.53 & 7.54 & 300 & 24.66 & On & $2.99 \times 10^7$ \\ \hline
5\labeltext{5}{case4_M8_nr} & 669 & 21.2 & 4.53 & 7.54 & 300 & 24.66 & Off & $2.98 \times 10^7$\\ \hline
\end{tabular}
\label{table:cases}
\end{table*}


\section{Flow configurations \label{sec:flow_config}}

A schematic of the computational domain is provided in Fig.\ \ref{fig:case_schematic}. The domain extends 30.4, 16, and 19.2 mm in the $x$-, $y$-, and $z$-directions, respectively. The injector diameter ($d$) is 0.4064 mm for all cases, so these grid dimensions equate to roughly 75, 39, and 47 jet diameters. A supersonic inflow with fixed pressure, temperature, and velocity is prescribed on the upstream ($-x$) face, while zero-gradient outflows are prescribed on the top ($+y$) and downstream ($+x$) faces. The transverse ($\pm z$) faces are adiabatic slip walls, while the bottom ($-y$) face is no-slip and adiabatic. The base grid has 152 $\times$ 80 $\times$ 96 cells, yielding a uniform resolution of 200 $\mu$m. Three AMR levels are used, each of which divides the cell size along all directions by a factor of two. As such, the minimum grid resolution in all directions is 25 $\mu$m. Tagging for refinement is performed using the liquid volume fraction value ($\alpha_l$) and the gradients of pressure ($\nabla p$), temperature ($\nabla T$), and liquid volume fraction ($\nabla \alpha_l$). Fixed refinement regions were added below $y/d = 12.3$ and in the near-injector region ($-7.4 \leq x/d \leq 7.4$; $y/d \leq 2.5$; $-5 \leq z/d \leq 5$). The former refined the entirety of roughly the bottom third of the domain with at least 1 AMR level, while the latter refined the region near the injector with the maximum AMR level. The resulting average numbers of cells, along with the key flow parameters, are provided in Table \ref{table:cases}.

Liquid n-dodecane is injected perpendicular to the crossflow through a boundary patch in the bottom ($-y$) domain face. The injection velocity is computed by
\begin{equation}
    u_{inj} = \sqrt{\frac{2(p_{plenum} - p_{int})}{\rho_{plenum}}}
    \label{eqn:u_inj}
\end{equation}
where $p_{int}$ is the static pressure at the internal cell neighboring each injection boundary face. If $p_{int}$ locally exceeds $p_{plenum}$, the boundary face instead becomes a no-slip wall. This condition was typically not observed in the simulations, so injection was sustained over the entire boundary patch during the simulations. This yielded the average momentum ratios $J$ listed in Table \ref{table:cases}.

Five cases were simulated in total, all with a freestream Mach number of roughly 4.5 and static pressure of 21.5 kPa. Cases \ref{case2} and \ref{case2_M8} utilize a lower plenum pressure, yielding $J \sim 12$, while cases \ref{case4}, \ref{case4_M8}, and \ref{case4_M8_nr} use a higher plenum pressure, yielding $J \sim 24$. Cases \ref{case2} and \ref{case4} use a lower freestream static temperature of $\sim$250 K, while cases \ref{case2_M8}-\ref{case4_M8_nr} use an elevated temperature of $\sim$685 K. Because the freestream Mach number and momentum ratio are held constant, this increase in freestream temperature corresponds to an increase in freestream velocity. In experiments conducted at the University of Central Florida \cite{ucf_jicf_personal}, configurations similar to cases \ref{case2}-\ref{case4} were found to exhibit little to no chemical reactions. As such, these cases were predominantly run with the chemistry reactions disabled in order to save on computational cost. Short run durations with the reactions turned on yielded very little heat release in these cases, confirming the experimental result. The chemical reactions were turned on for the duration of cases \ref{case2_M8} and \ref{case4_M8}, as the elevated flow enthalpy was expected to lead to more substantial heat release. In order to assess the effects of this heat release, case \ref{case4_M8_nr} was run with the same conditions as case \ref{case4_M8}, but with the reactions turned off for the entire simulation duration.


\section{Results \label{sec:results}}


\subsection{General flow features}

\begin{figure}[!h]
     \centering
     \includegraphics[width=\textwidth]{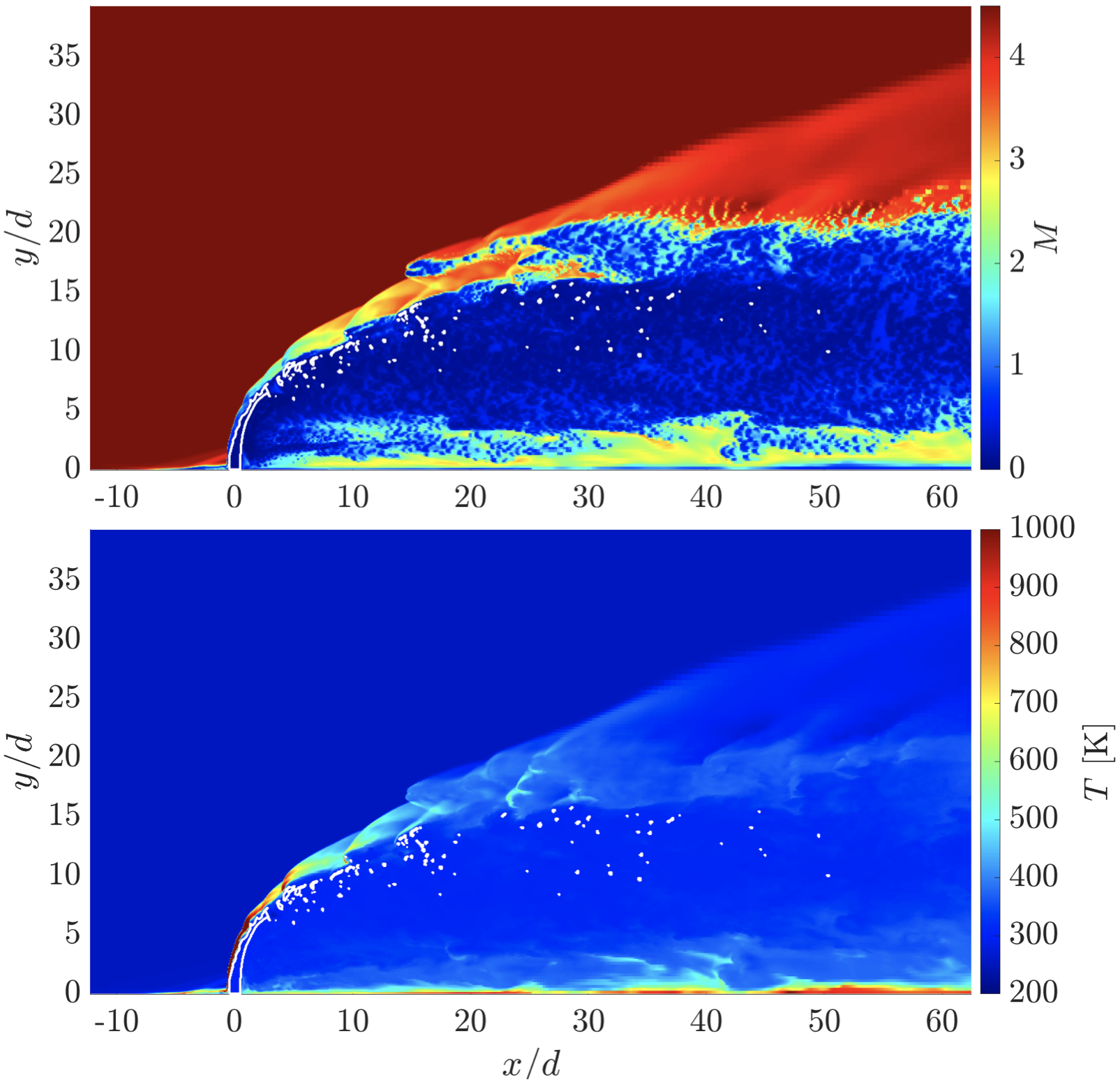}
    \caption{Instantaneous Mach number (top) and temperature (bottom) contours along the $z$-midplane in case \ref{case2}. $\alpha_l=1$ isocontour marked in white.}
    \label{fig:mach_T_case2}
\end{figure}

\begin{figure}[!h]
     \centering
     \includegraphics[width=\textwidth]{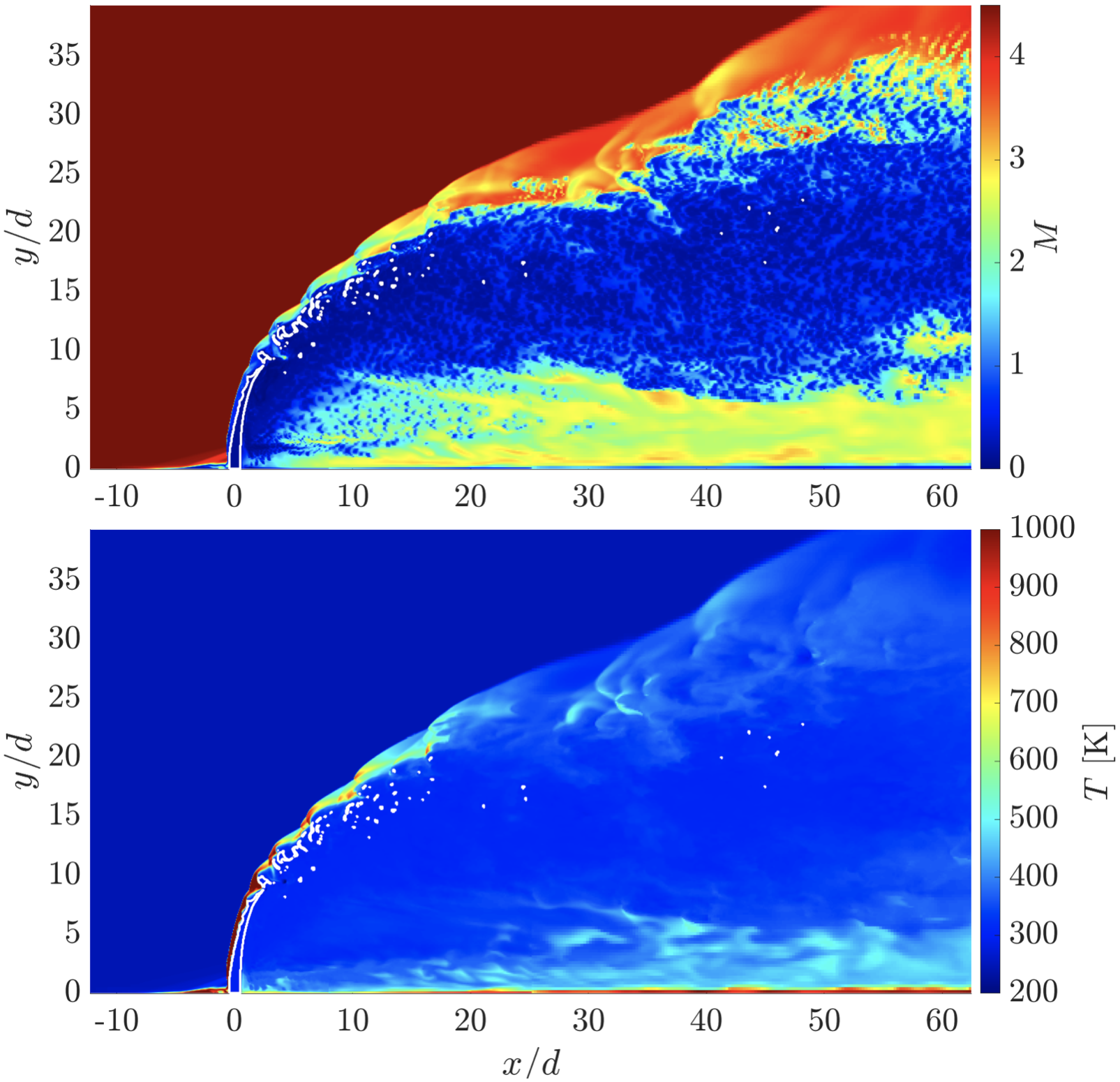}
    \caption{Instantaneous Mach number (top) and temperature (bottom) contours along the $z$-midplane in case \ref{case4}. $\alpha_l=1$ isocontour marked in white.}
    \label{fig:mach_T_case4}
\end{figure}

Figures \ref{fig:mach_T_case2} and \ref{fig:mach_T_case4} show instantaneous contours of Mach number and temperature along the $z$-midplane in cases \ref{case2} and \ref{case4}. In both cases, similar features can be observed. The separation shock upstream of injection impinges upon the bow shock, creating instabilities along the bow shock. These instabilities grow as they travel up the bow shock, generating turbulence in its wake and facilitating the liquid atomization at the top of the liquid injection column. The region in the jet wake is dominated by subsonic flow, with the exception of the region near the bottom wall—$y/d<5$ in case \ref{case2}; $y/d<10$ in case \ref{case4}—where the comparatively narrow liquid column creates a smaller obstruction to the freestream than the spreading and atomizing liquid jet at greater heights. This effect can also be seen in Fig.\ \ref{fig:case_schematic}. Elevated temperatures are only observed along the bow shock, in the recirculation zone upstream of injection, and in the boundary layer downstream of injection. In the majority of the jet wake, the temperature is only $\sim$150 K higher than the freestream. This is because the enthalpy of the freestream is largely expended by the warming and evaporation of the liquid, leaving little energy available to warm the remaining gas. This helps to explain why negligible chemical reactions were observed in the complementary experiments \cite{ucf_jicf_personal} and in the short durations where these cases were run with chemical reactions enabled—the jet wake is simply too cold for any evaporated fuel to react.

\begin{figure}[!h]
     \centering
     \includegraphics[width=\textwidth]{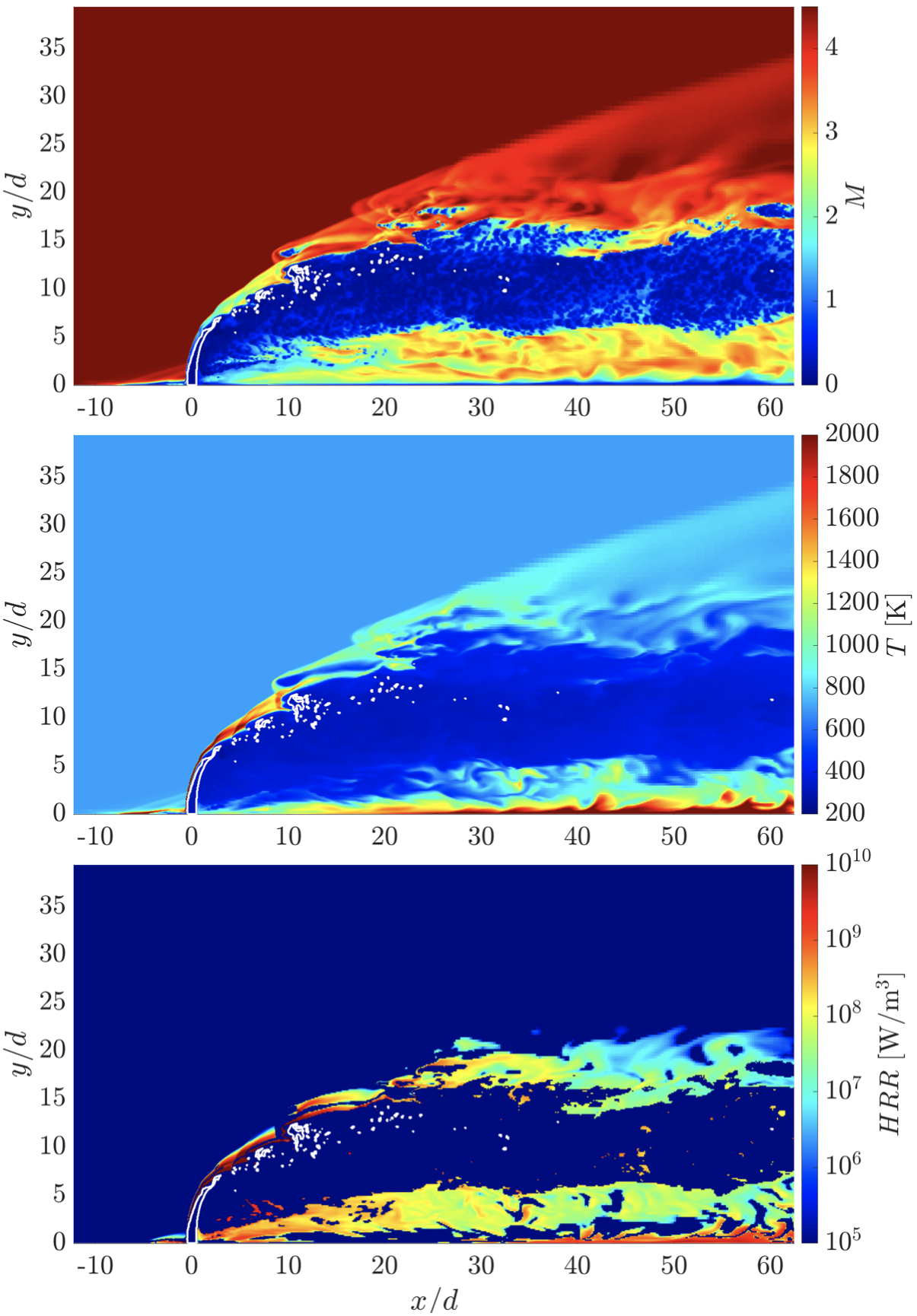}
    \caption{Instantaneous Mach number (top), temperature (middle), and volumetric heat release rate (bottom) contours along the $z$-midplane in case \ref{case2_M8}. $\alpha_l=1$ isocontour marked in white.}
    \label{fig:mach_T_HR_case2_M8}
\end{figure}

\begin{figure}[!h]
     \centering
     \includegraphics[width=\textwidth]{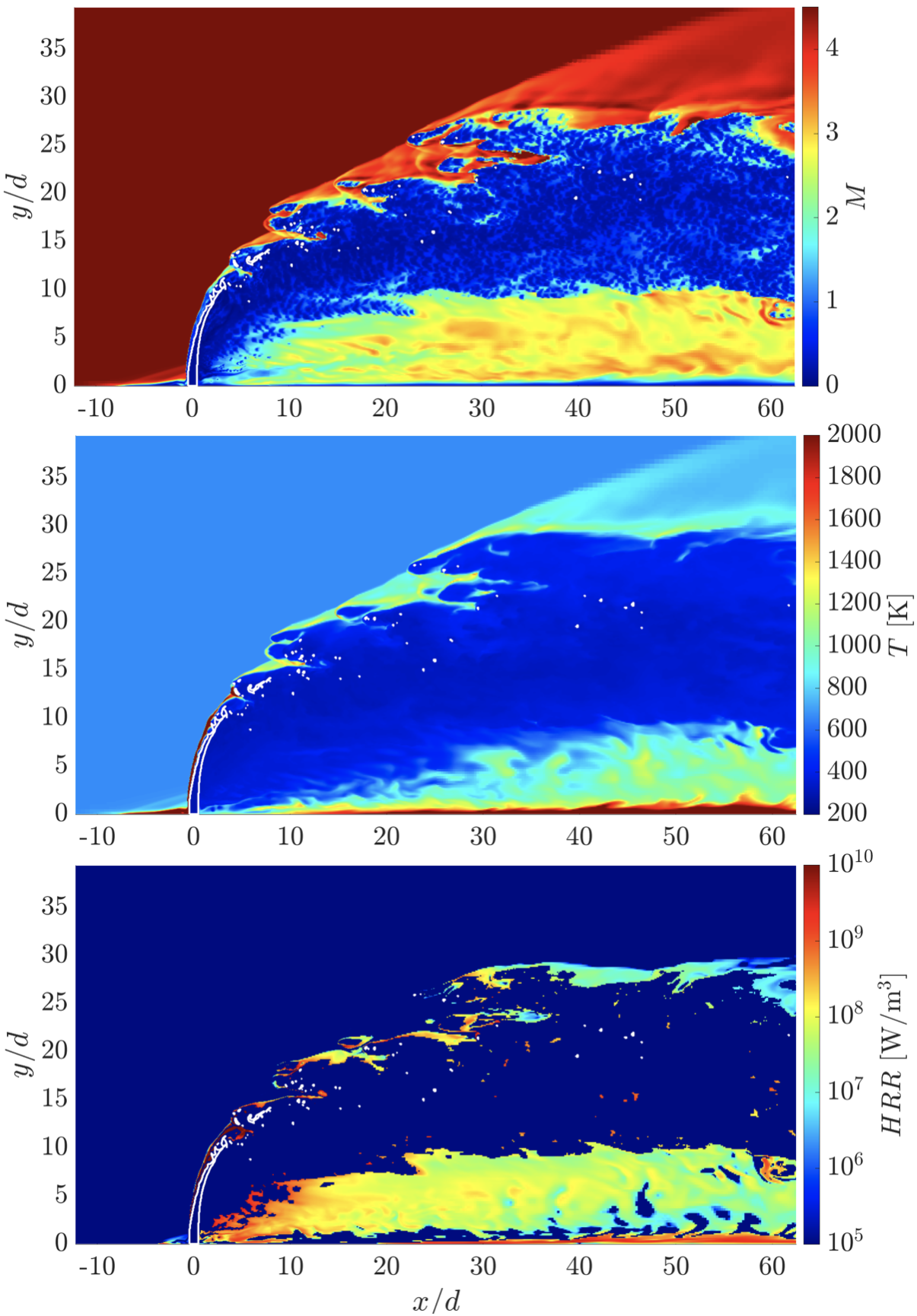}
    \caption{Instantaneous Mach number (top), temperature (middle), and volumetric heat release rate (bottom) contours along the $z$-midplane in case \ref{case4_M8}. $\alpha_l=1$ isocontour marked in white.}
    \label{fig:mach_T_HR_case4_M8}
\end{figure}

This is contrasted against cases \ref{case2_M8} and \ref{case4_M8}, which utilized higher freestream temperatures. Instantaneous $z$-midplane contours of Mach number, temperature, and volumetric heat release rate for these cases are plotted in Figs.\ \ref{fig:mach_T_HR_case2_M8} and \ref{fig:mach_T_HR_case4_M8}. Here, many of the same features can be seen as in cases \ref{case2} and \ref{case4}. The separation shock triggers instabilities along the bow shock, which facilitate turbulence and liquid atomization at the top of the liquid column. The jet wake is again largely subsonic above the height of the liquid column and supersonic below this point ($y/d\sim5$ in case \ref{case2_M8}; $y/d\sim10$ in case \ref{case4_M8}. However, the elevated freestream leads to higher temperatures along the bow shock, in the upstream recirculation zone, and in the near-wall region immediately downstream of injection ($x/d<10$, $y/d<5$). Heat release is correspondingly observed in these three regions. Interestingly, heat release is largely absent injector wake at the injection height. This is due to low temperatures here induced by the evaporation of atomized fuel droplets. Instead, the strongest heat release is predominately observed between the bow shock and the upper half of the liquid column. Here, injected fuel is able to quickly evaporate and react due to the high pressures and temperatures induced by the bow shock. Progressing upward, the atomization at the top of the liquid column and sustained strength of the bow shock facilitate high heat release in the oblique shock above the injection height.

In the upstream recirculation zone and counter-rotating vortices on the $\pm z$ sides of the injection orifice, the residence time for the fuel is increased. This gives sufficient time for it to evaporate and begin to react, leading to the formation of intermediate products such as C$_2$H$_4$, C$_3$H$_6$, and OH in these near-wall regions. These reactions continue to progress within the low-velocity regions on the leeward side of the jet, yielding the elevated heat release observable at $0 < x/d < 10$, $y/d<5$ and along the bottom wall for $x/d>25$. As will be discussed further in Sec.\ \ref{sec:combustion_efficiency}, the increased residence times in the leeward boundary layer facilitate the formation of final products, such as CO, CO$_2$, and H$_2$O, along the bottom wall for $x/d>20$. Altogether, the reduction in evaporation time and ignition delay enabled by the elevated freestream temperature allow more substantial chemical reactions to occur in cases \ref{case2_M8} and \ref{case4_M8}.


\subsection{Jet penetration and trajectory}

The jet penetration is a key quantity of interest for JISC configurations, as it is one of the driving factors in facilitating the mixing of the injected fluid with the freestream. Gaseous JISC studies have found the jet penetration to be a function of the jet-to-crossflow momentum ratio ($J$), the entrance boundary layer thickness ($\delta$), and the molecular weight of the injected gas ($M$) \cite{Yakar2006, portz2005transverse, portz2006penetration, mcclinton1974effect, sharma2024_scitech, fries2021turbulent}. Liquid JISC studies, on the other hand, have primarily focused on the influence of the momentum ratio \cite{perurena2009experimental, ghenai2009, lin2004_pdpa, yates1972liquid}, although Sharma et al.\ \cite{sharma2025_aiaa} did observe in increase in liquid jet penetration with an increased entrance boundary layer thickness, as is typically observed with gaseous injection.

\begin{figure}[!h]
    \centering
    \includegraphics[width=\textwidth]{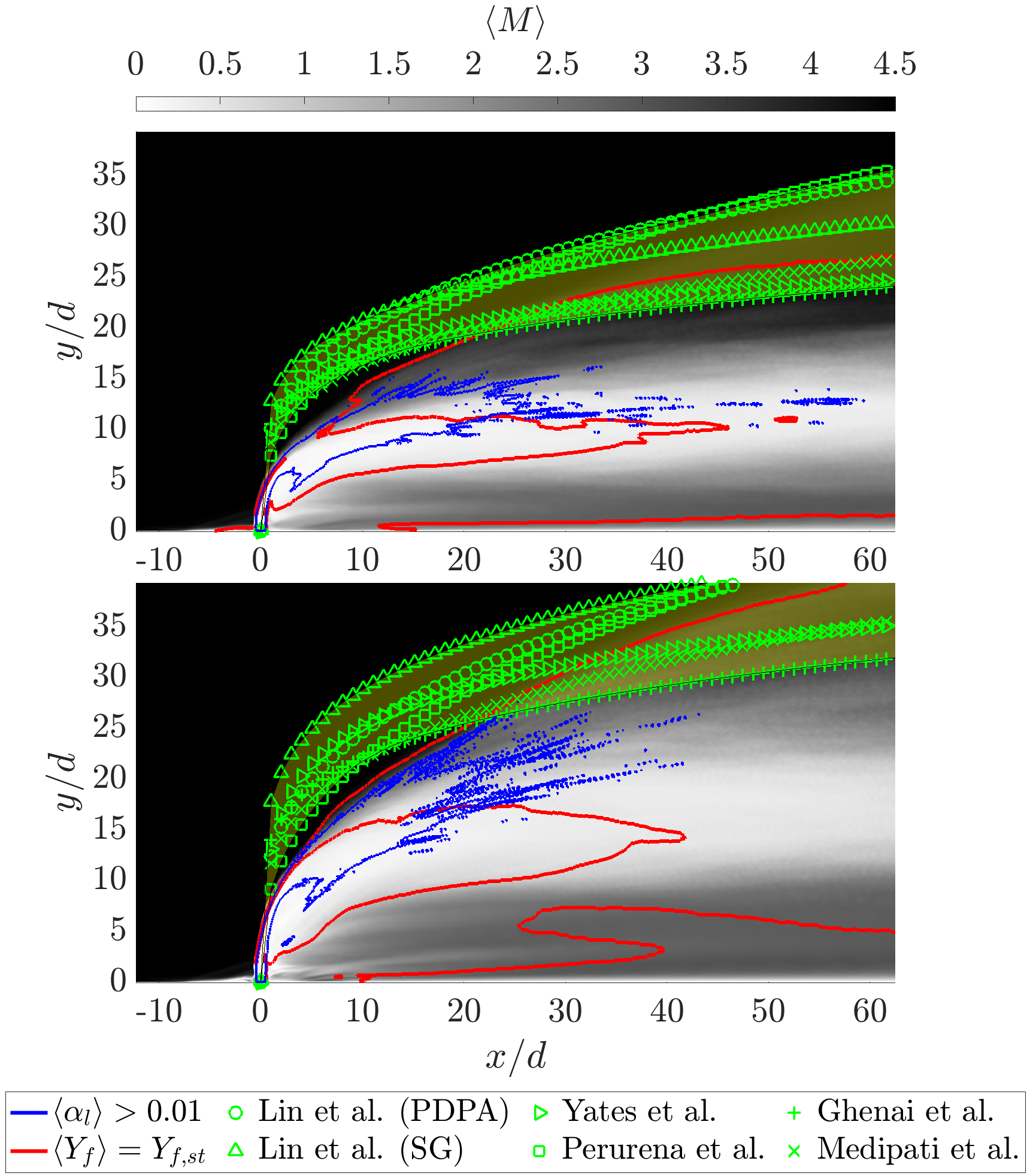}
    \caption{Time-averaged Mach number contours along the $z$-midplane in cases \ref{case2_M8} (top) and \ref{case4_M8} (bottom). Penetration correlations plotted in green, with span of correlations plotted in yellow.}
    \label{fig:mach_avg_traj}
\end{figure}

Various methods and metrics have been used to ascertain the jet penetration height in experiments, including time-averaged Schlieren and shadowgraph images \cite{portz2005transverse, portz2006penetration, lin2002penetration}, fluid concentration \cite{sun2018formation, mcclinton1974effect, rogers1971study, povinelli1971correlation, fries2021turbulent}, phase Doppler particle analysis (PDPA) \cite{lin2004_pdpa, lin2017_pdpa}, and, in reacting scenarios, OH-PLIF \cite{gamba2015_jisc, rothstein1992study}. From these height measurements, empirical correlations are then derived for the jet penetration height $y/d$ as a function of downstream distance $x/d$ and the flow parameters ($J$, $\delta$, etc.). In order to compare the present cases against existing liquid JISC literature, Fig.\ \ref{fig:mach_avg_traj} shows the time-averaged Mach number profiles along the $z$-midplane in cases \ref{case2_M8} and \ref{case4_M8}. The profiles for the remaining cases were similar and are thus omitted here for brevity. Figure \ref{fig:mach_avg_traj} also shows isocontours for the time-averaged liquid volume fraction $\langle \alpha_l \rangle$ and the mass fraction of gaseous fuel $\langle Y_f \rangle$, which give insight into their respective mean trajectories. Alongside the simulation results are empirical liquid jet penetration correlations from several experimental studies \cite{lin2004_pdpa, lin2002penetration, yates1972liquid, perurena2009experimental, ghenai2009, medipati2023liquid}, all of which were functions of the momentum ratio ($J$) only.

Figure \ref{fig:mach_avg_traj} shows that the current simulations tend to underpredict the liquid penetration heights compared to the experimental empirical correlations. While the liquid column extends to $y/d \sim 6$ in case \ref{case2_M8} and $y/d \sim 9$ in case \ref{case4_M8}, the spray of atomized droplets—marked by the $\langle \alpha_l \rangle > 0.01$ contour—extends farther up into the low-Mach plume. Both cases exhibit discrepancies for $x/d<20$, but the spray trajectory in case \ref{case4_M8} does approach the lower bound of the empirical correlations farther downstream. Due to advection of evaporated fuel along the bow and oblique shock trajectories, the isoline for the stoichiometric fuel mass fraction extends to the top of the low Mach plume, such that it lies within the bounds of the empirical correlations for $x/d>25$.

\begin{figure*}[!b]
     \centering
     \begin{subfigure}[b]{0.35\textwidth}
         \centering
         \includegraphics[width=\textwidth]{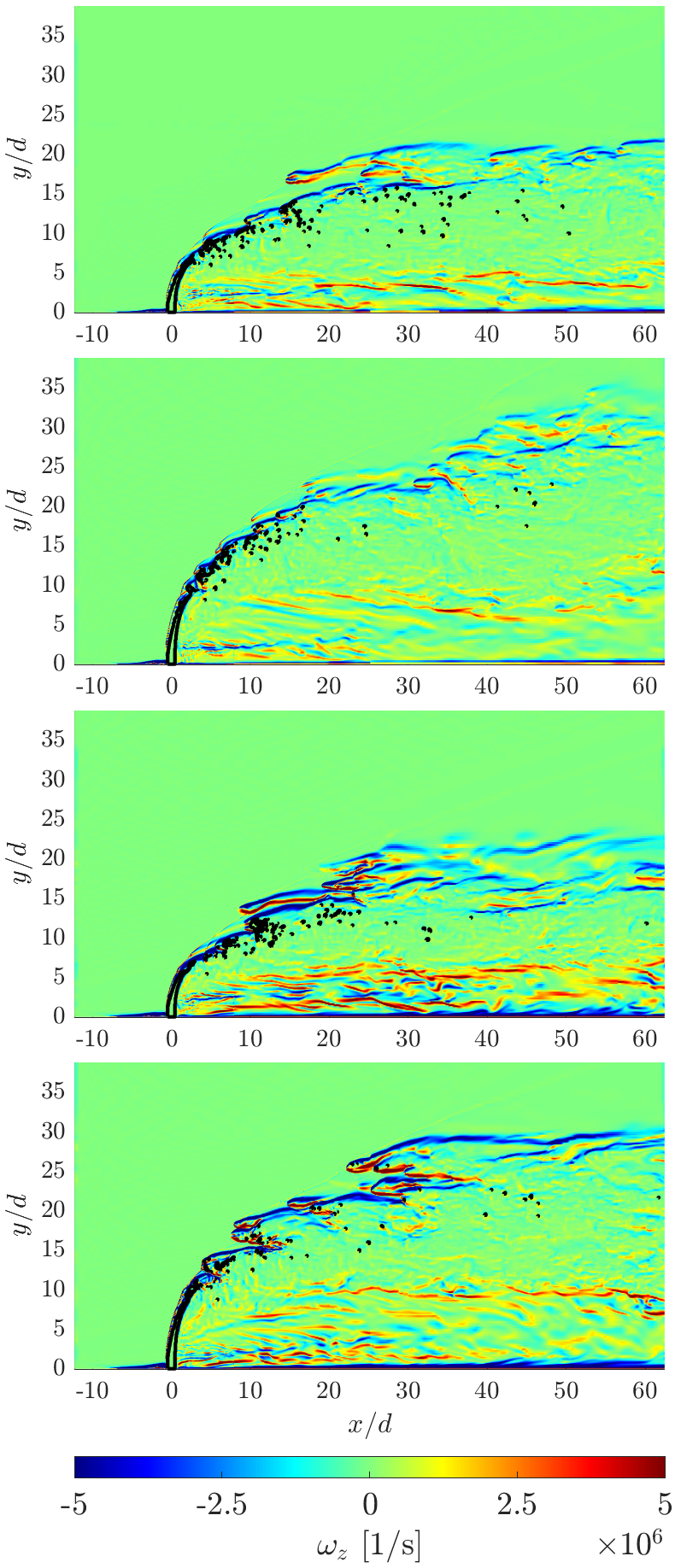}
     \end{subfigure}
     \hspace{5mm}
     \begin{subfigure}[b]{0.435\textwidth}
         \centering
         \includegraphics[width=\textwidth]{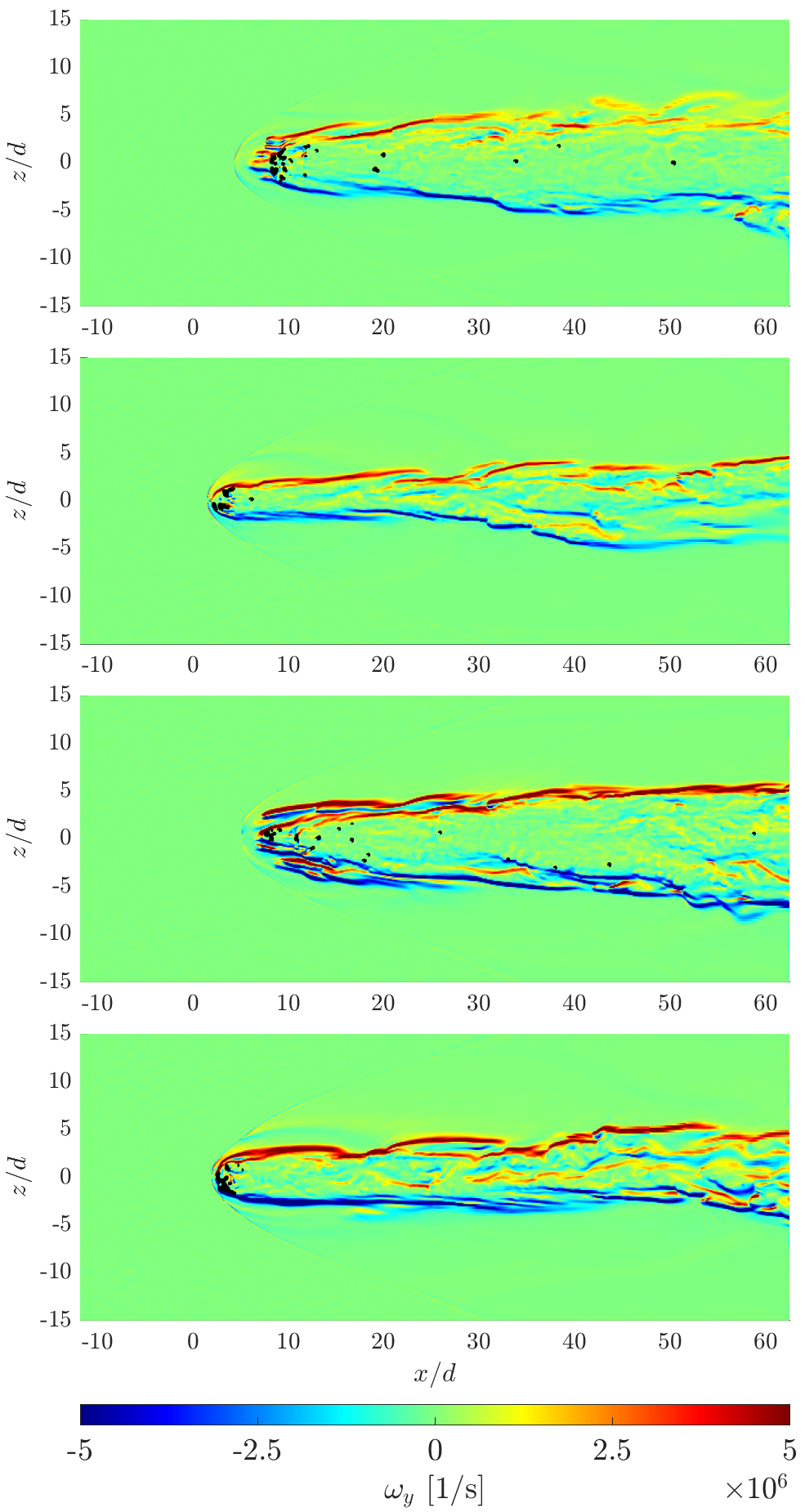}
     \end{subfigure}
    \caption{From top to bottom, instantaneous vorticity contours in cases \ref{case2}-\ref{case4_M8}. (Left) $z$-component of vorticity along the $z$-midplane. (Right) $y$-component of vorticity along a plane at a height of $y/d=10$. $\alpha_l=1$ isocontours marked in black.}
    \label{fig:vort_yz_inst}
\end{figure*}

\begin{figure*}[!b]
     \centering
     \begin{subfigure}[b]{\textwidth}
         \centering
         \includegraphics[width=\textwidth]{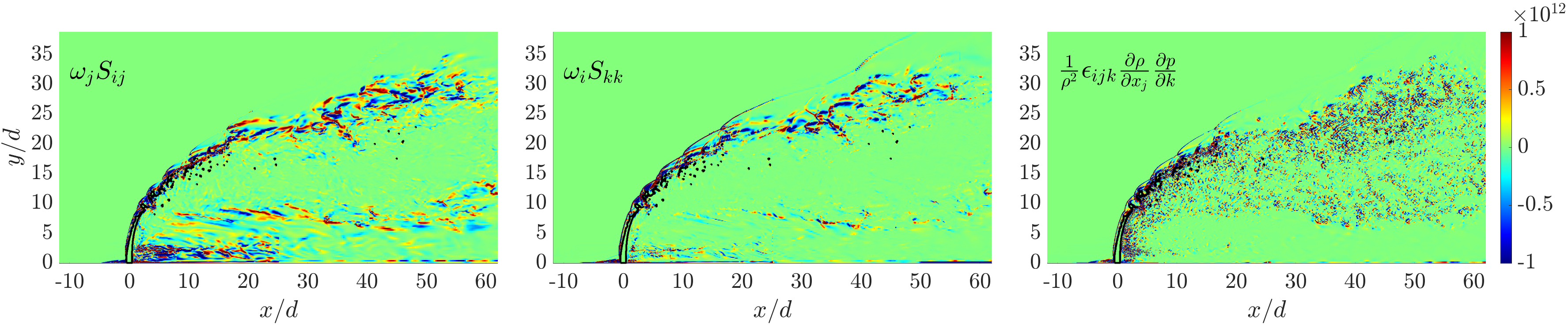}
     \end{subfigure}
     \\
     \begin{subfigure}[b]{\textwidth}
         \centering
         \includegraphics[width=\textwidth]{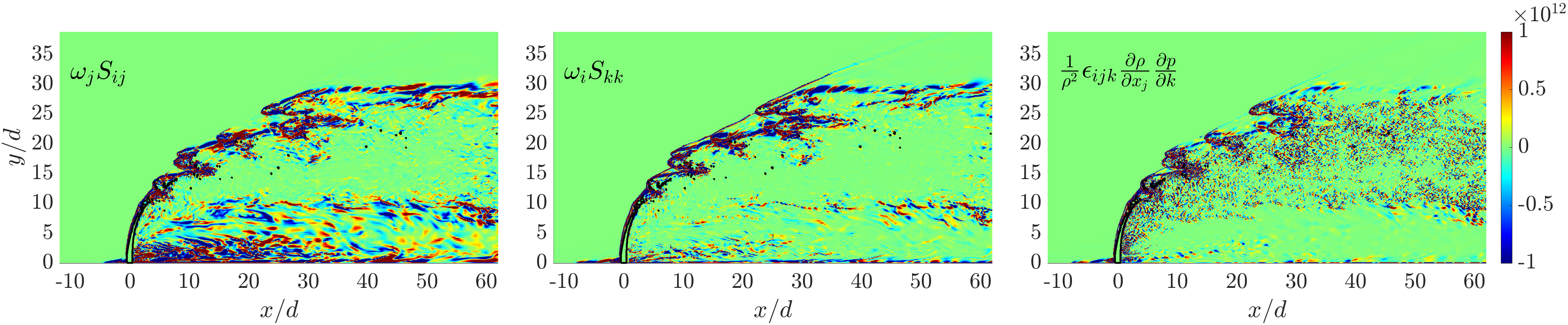}
     \end{subfigure}
     \\
     \begin{subfigure}[b]{\textwidth}
         \centering
         \includegraphics[width=\textwidth]{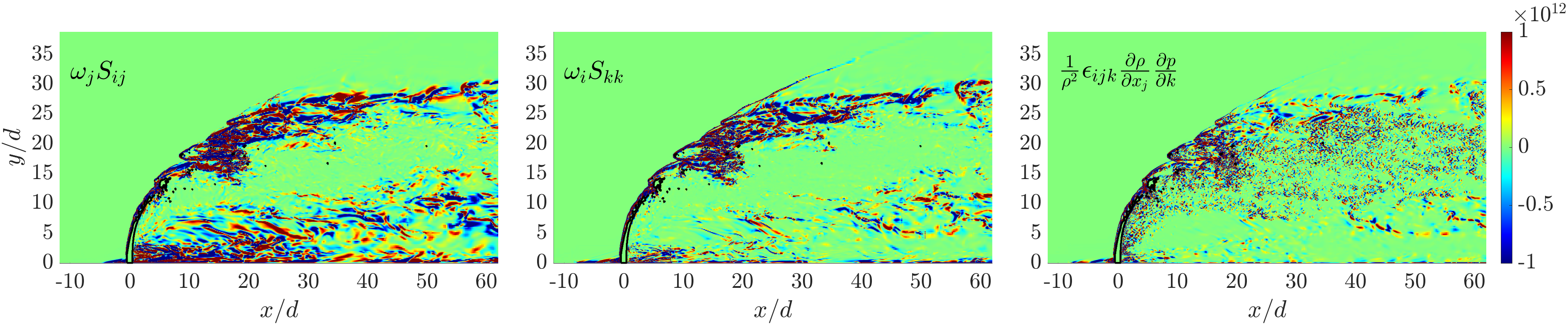}
     \end{subfigure}
    \caption{Instantaneous contours of components of the vorticity transport equation along the $z$-midplane in cases \ref{case4} (top), \ref{case4_M8} (middle), and \ref{case4_M8_nr}. $\alpha_l=1$ isocontours marked in black.}
    \label{fig:vort_comp_z_inst}
\end{figure*}

It is hypothesized that the underprediction in jet penetration is related to the thickness of the boundary layer upstream of injection. In the experiments of Medipati et al.\ \cite{medipati2023liquid} and Lin et al.\ \cite{lin2002penetration}, for instance, the boundary layer thicknesses upstream of injection were respectively measured as 8.8 and 12.4 times the orifice diameter. These are similar to the penetration heights predicted by their respective correlations near the injection point. In the current simulations, the freestream conditions are prescribed uniformly across the inflow boundary, and the short distance upstream of injection ($x/d \sim 12$) gives little opportunity for a boundary layer to develop (as seen in Figs.\ \ref{fig:mach_T_case2}-\ref{fig:mach_T_HR_case4_M8}). Previous simulations have noted an increase liquid jet penetration when the upstream boundary layer thickness is increased \cite{sharma2025_aiaa}, as this increases the effective jet momentum ratio by decreasing the momentum of the freestream flow impinging upon the jet. It is expected that similar behavior would be observed here if a boundary layer was explicitly prescribed at the inflow.


\subsection{Mixing analysis}


A major advantage of transverse fuel injection into supersonic crossflow is increased mixing compared to other injection configurations, such as angled injection and strut-based injection \cite{gruber1995_jisc}. The resultant high-speed interaction of fuel and crossflow produces spatiotemporal complexities that affect both the kinematic and dynamic properties of turbulence. The kinematic properties can be analyzed through the use of velocity gradient quantities (e.g., vorticity) and the second-order statistics of velocity, including turbulent kinetic energy (TKE) and Reynolds stresses. The dynamic properties, on the other hand, pertain to the underlying physical mechanisms that influence the structure and movement of turbulent flow structures \cite{steinberg2021structure}. In this section, these properties are examined to provide insight into the mixing processes and their underlying mechanisms.

One consequence of curved shocks and chemical heat release in JISC configurations is the generation and transport of vorticity, which facilitates the development of multiscale turbulence \cite{sharma2024_proci}. Figure~\ref{fig:vort_yz_inst} shows the vorticity components across the $z$-midplane and wall-normal plane $y/d = 10$ for cases \ref{case2}-\ref{case4_M8}. In all cases, the $z$-component of vorticity on the $z$-midplane indicates that the windward shear layer formed by the interaction of the jet with the crossflow is the main source of vorticity generation. This can be seen in the $y$-component of vorticity as well, which is primarily concentrated along the transverse shear layers on either side of the injection point. The increased vorticity in cases \ref{case2_M8} and \ref{case4_M8} is likely due to the higher freestream velocity, which subsequently generates larger velocity gradients when interacting with the liquid jet and atomized droplets. In order to understand the physical processes that lead to vorticity generation, the vorticity transport equation is considered. The transport equation for the $i$-th directional vorticity component can be written as \cite{steinberg2021structure}:

\begin{equation}
    \frac{D \omega_i}{Dt} = \omega_j S_{ij} - \omega_i S_{kk} + \frac{1}{\rho^2} \epsilon_{ijk} \frac{\partial \rho}{\partial x_j} \frac{\partial p}{\partial x_k} + \epsilon_{ijk} \frac{\partial}{\partial x_j} \left( \frac{1}{\rho} \frac{\partial \tau_{kl}}{\partial x_l} \right)
    \label{eqn:vorticity}
\end{equation}
where the four terms on the right-hand side respectively denote the vortex stretching associated with the strain rate $S_{ij}$, the effect of dilatation due to compressibility, the baroclinic torque generated from the misalignment of pressure and density gradients, and the viscous transport term. The latter is taken to be negligible for the high-speed cases considered here. 

Figure \ref{fig:vort_comp_z_inst} shows the components of the vorticity transport equation for cases \ref{case4}, \ref{case4_M8}, and \ref{case4_M8_nr}. Only these three cases are shown for brevity, as cases \ref{case2} and \ref{case2_M8} show similar trends. The incoming crossflow in all the cases is not turbulent, meaning that the jet structures—namely, the curved shocks and windward shear layer—are the primary generators of vorticity. The left plots in Fig.\ \ref{fig:vort_comp_z_inst} show that the vortex stretching associated with the strain rate is the primary mechanism for amplification and redistribution of vorticity in all three cases. On the leeward side, the jet breakup and subsequent droplet formation in the low pressure region facilitate vorticity generation as well. For all cases, the formation of ligaments and droplets along the jet core, as seen in Figs.\ \ref{fig:mach_T_case2}-\ref{fig:mach_T_HR_case4_M8}, produces large density gradients. The misalignment in the pressure and density gradients in the jet wake results in baroclinic torque generation, as seen in the right plots in Fig.\ \ref{fig:vort_comp_z_inst}. This leads to sustained production of vorticity in this region. The dilatation component, reflecting compressibility effects, is most prevalent in the windward jet shear layer, where its production peaks as detached liquid clumps generate small shocks. Interestingly, the reacting and non-reacting cases with the same freestream conditions (\ref{case4_M8} and \ref{case4_M8_nr}) show similar profiles for all three terms along the windward bow shock and shear layer, as well as in the jet wake. Previous work on premixed jet flames has shown that chemical heat release can lead to amplification shear and vortex stretching \cite{coriton2016experimental}. However, in these complementary cases, the vortex stretching term is of similar magnitude along the bow shock and in the jet wake—the two primary heat release zones shown in Fig.\ \ref{fig:mach_T_HR_case4_M8}. Similar to previous simulations of gaseous JISC \cite{sharma2024_proci}, this suggests that the heat release is too weak to generate a substantial amount of vorticity. Instead, for the current set of cases, vortex stretching within the shear layers is the predominant generator of vorticity for the windward side and the near-wall reacting region in the jet wake. The contours along the $y/d = 10$ plane in Fig.\ \ref{fig:vort_yz_inst} show similar trends, in which the shear layer is the main source of vorticity in all cases.

\begin{figure*}[!h]
    \centering
    \includegraphics[width=\textwidth]{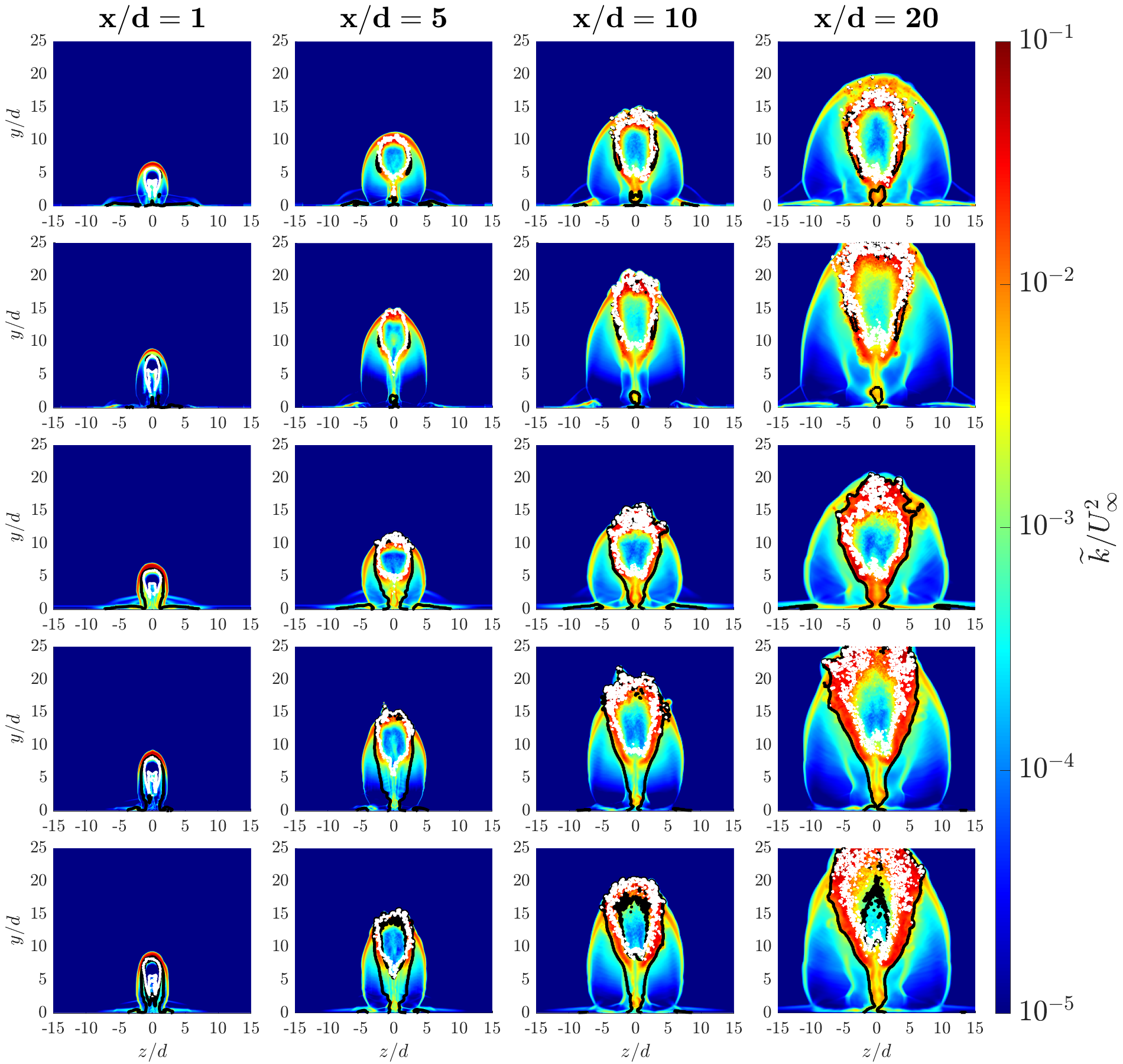}
    \caption{Favre-averaged turbulent kinetic energy along different $x/d$ planes for all cases, with row corresponding to case number. $\alpha_l>0.01$ isocontours colored in white. $Z_{mix,g} = Z_{st}$ isocontours colored in black.}
    \label{fig:TKE_favre_norm}
\end{figure*}

To quantify mixing effects, the Favre-averaged turbulent kinetic energy (TKE) is computed at different $x/d$ locations. The time average is replaced by the Favre-averaged definitions here to account for the spatiotemporal variations in density. The Favre-averaged fields are computed as
\begin{equation}
    \psi = \frac{\overline{\rho \psi}}{\overline{\rho}}
    \label{eqn:favre_avg}
\end{equation}
where $\psi$ represents the quantity of interest and $\overline{\rho}$ is the time-averaged density. The Favre-averaged TKE is given as
\begin{equation}
    \widetilde{k} = \frac{1}{2} \left( \widetilde{u'u'} + \widetilde{v'v'} + \widetilde{w'w'} \right)
    \label{eqn:tke_favre_avg}
\end{equation}
where $u'$ is the fluctuating component in Favre averaging. Figure~\ref{fig:TKE_favre_norm} shows the Favre-averaged TKE computed at different $x/d$ planes for cases \ref{case2}-\ref{case4_M8}. In each, the TKE has been normalized by the square of the freestream velocity, $U_\infty$.

Consistent with the observations made for vorticity generation and transport, the TKE intensity is highest in the jet shear layers. In the near-field of the jet, at $x/d = 1$ in Fig.\ \ref{fig:TKE_favre_norm}, the liquid jet column remains intact, with no significant breakup or droplet formation. Consequently, the TKE intensity is lower in this region, with some fluctuations caused by the movement of the bow shock. Farther downstream in the jet wake, the TKE intensity increases as a result of the primary and secondary breakup of the liquid jet column. The formation of ligaments and droplets sheared from the column ($x/d \geq 5$), shown through isocontours of the liquid volume fraction $\alpha_l > 0.01$, leads to the generation of baroclinic torque, which contributes to increased vorticity generation and TKE intensity. This is further evident from the high intensity region in Fig.\ \ref{fig:TKE_favre_norm} located between the outer boundary of the jet core (white isolines) and the isolines for the stoichiometric mixture fraction in the gas phase (black). This reinforces the earlier observation that the shear layer drives much of the turbulence generation in these cases, consistent with prior studies on both gaseous and liquid jets \cite{sharma2024_proci, cao2020confinement, sharma2025_aiaa}. Meanwhile, the similarities between cases \ref{case4_M8} and \ref{case4_M8_nr} show that the presence of chemical reactions has limited impact on vorticity generation and subsequent TKE intensity.


\subsection{Combustion efficiency \label{sec:combustion_efficiency}}

\begin{figure}[!h]
    \centering
    \includegraphics[width=\textwidth]{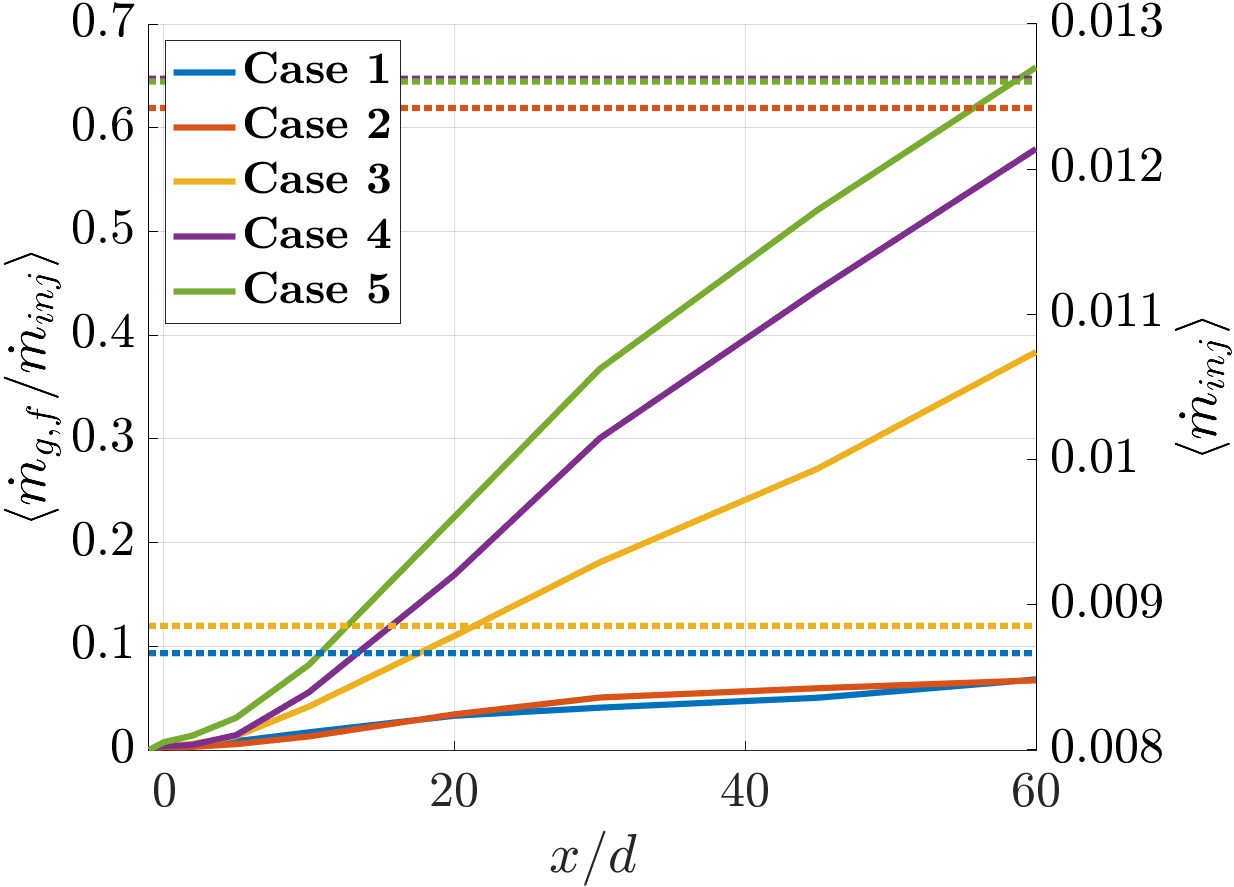}
    \caption{(Solid lines; left axis) Time-average of gaseous fuel mass flow rate normalized by injection mass flow rate at various $x/d$ planes. (Dashed lines; right axis) Time average of injection mass flow rate.}
    \label{fig:mdotx_Y_f__mdot_inj__timeavg}
\end{figure}

The combustion efficiency is primarily driven by the time and—via local flow velocity—length scales associated with evaporation and ignition of the fuel. The former is illustrated in Fig.\ \ref{fig:mdotx_Y_f__mdot_inj__timeavg}, which shows the mass flow rate of gaseous fuel versus the liquid injection flow rate at different $x/d$ planes. As such, the flow rate for the gaseous fuel is evaluated using the $x$-component of velocity. As may be expected, all cases exhibit a growing gaseous fuel flow rate over distance. This is indicative of the evaporation of liquid fuel in the jet wake and, in the reactive cases, illustrates that the conversion from liquid to gaseous fuel dominates the conversion of liquid fuel to product gases. The lower freestream temperatures in cases \ref{case2} and \ref{case4} lead to substantially lower evaporation rates, such that the gaseous fuel flow rate is roughly 7\% of the injected flow rate at $x/d=60$. Between cases \ref{case2_M8} and \ref{case4_M8}, there are growing discrepancies in gaseous fuel flow rate after $x/d=5$. The faster evaporation in cases \ref{case4_M8} and \ref{case4_M8_nr} is likely due to a combination of their stronger bow shocks—induced by their larger injection momentum ratio—and the higher mean temperatures in their jet wakes.

\begin{figure}[!h]
    \centering
    \includegraphics[width=\textwidth]{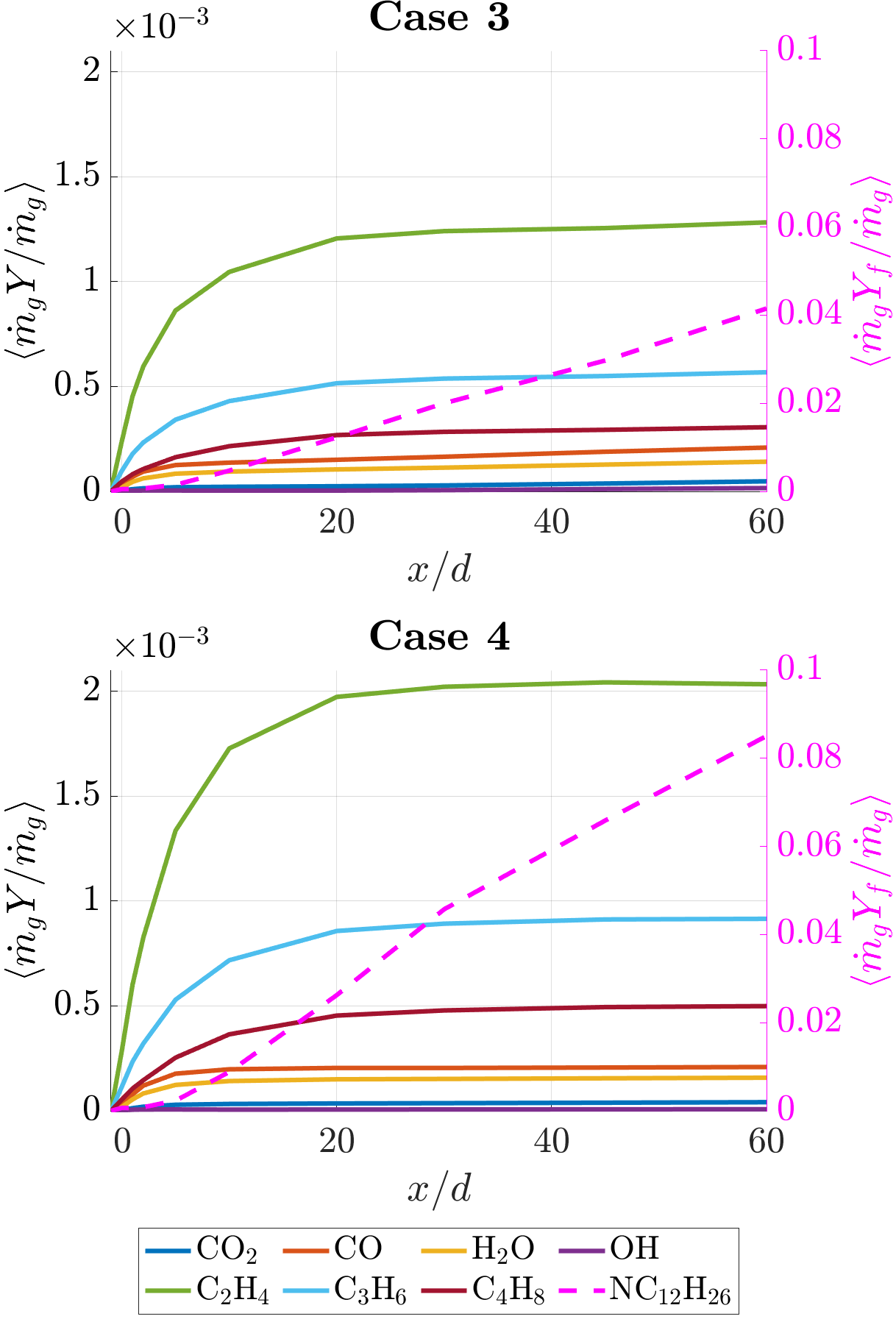}
    \caption{Time-average of gaseous species mass flow rates normalized by total gaseous mass flow rate at various $x/d$ planes. (Solid lines; left axes) Product gas species. (Dashed lines, right axes) Gaseous fuel.}
    \label{fig:mdotx_Y__mdotx_g__timeavg}
\end{figure}

\begin{figure}[!h]
    \centering
    \includegraphics[width=\textwidth]{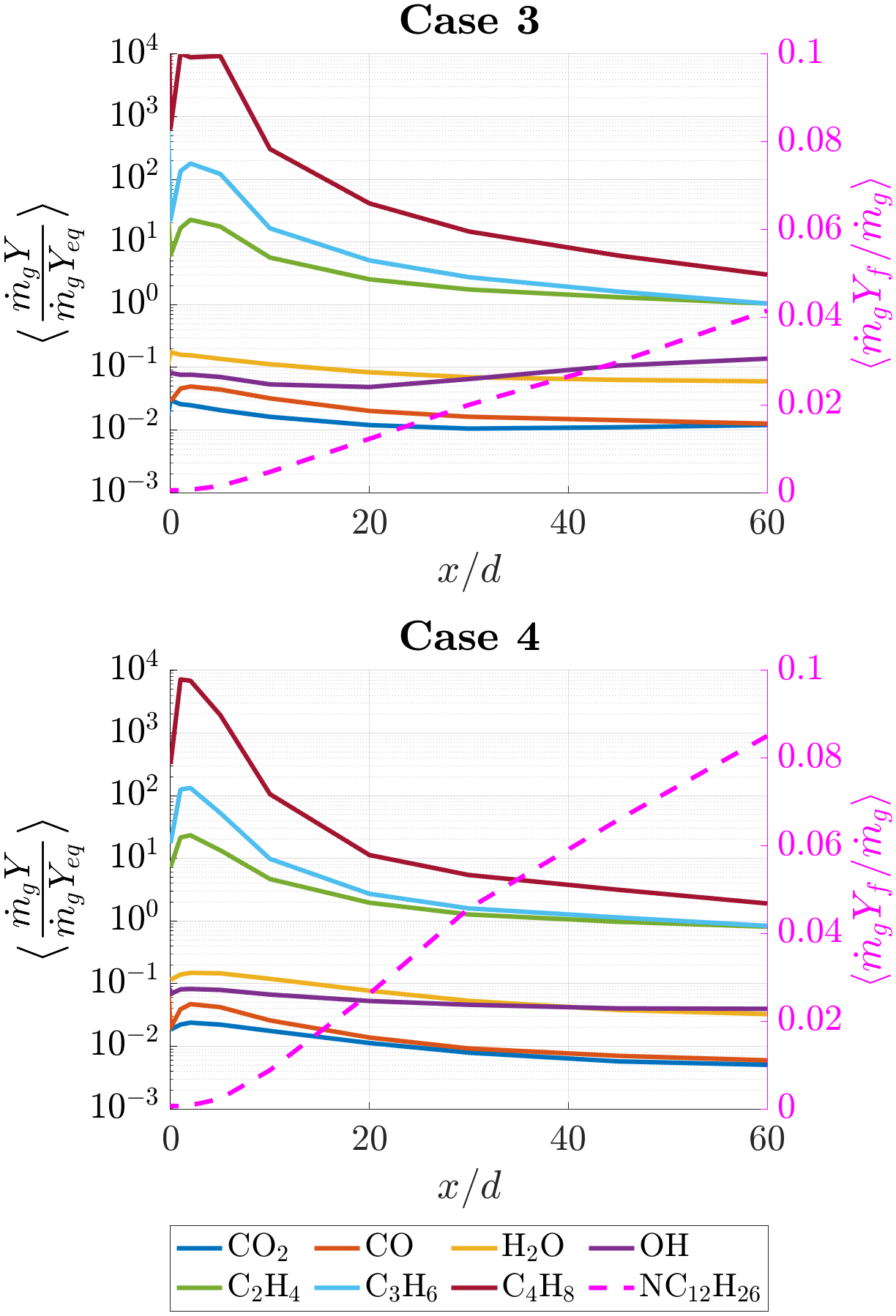}
    \caption{(Solid lines; left axes) Time-average of gaseous species mass flow rates normalized by mass flow rates of equilibrium gaseous species at various $x/d$ planes. (Dashed lines; right axes) Time-average of gaseous fuel mass flow rate normalized by total gaseous mass flow rate.}
    \label{fig:mdotx_Y___mdotx_Y_eq__timeavg}
\end{figure}

\begin{figure}[!h]
    \centering
    \includegraphics[width=\textwidth]{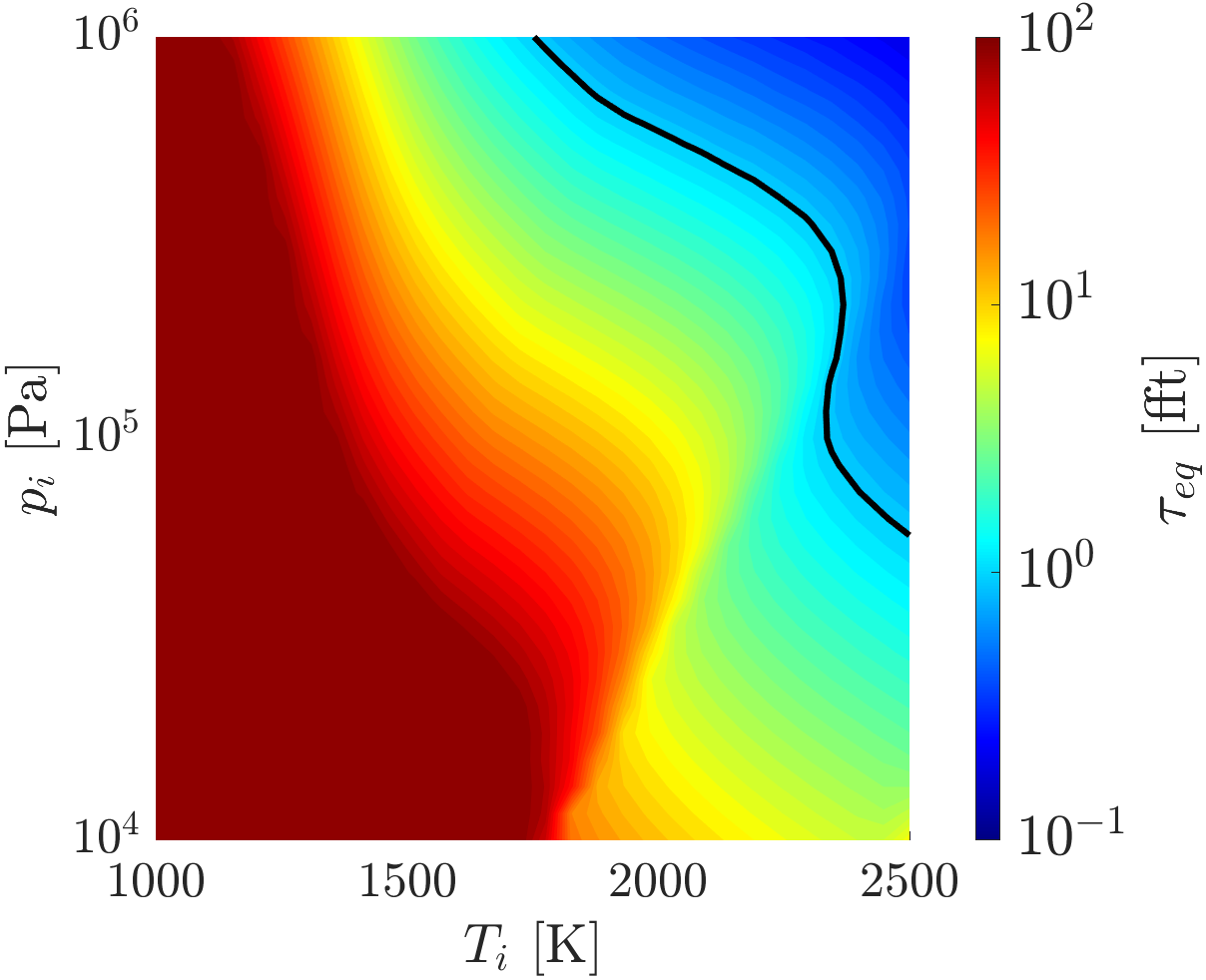}
    \caption{Time required to reach within 10\% of equilibrium composition in zero-dimensional constant-pressure reactor simulations with stoichiometric gaseous n-dodecane-air mixtures. $\tau_{eq} = 1$ fft isoline marked in black.}
    \label{fig:eq_delay}
\end{figure}

To examine the combustion dynamics in cases \ref{case2_M8} and \ref{case4_M8}, Fig.\ \ref{fig:mdotx_Y__mdotx_g__timeavg} shows the mass flow rates of gaseous fuel and selected product species normalized by the total gaseous mass flow rate at various distances from injection. Because case \ref{case4_M8} has a greater fueling rate, the flow rates of all plotted species are correspondingly larger. In both cases, the product gases are largely dominated by intermediate product species, such as C$_2$H$_4$, C$_3$H$_6$, and C$_4$H$_8$, which reach their maximum flow rates at around $x/d=30$. Final products, such as CO, CO$_2$, and H$_2$O, make up little of the gaseous flow rate by comparison. This suggests that the majority of the reactive flow field has not been given sufficient time to reach equilibrium prior to exiting the domain. This can be seen more directly in Fig.\ \ref{fig:mdotx_Y___mdotx_Y_eq__timeavg}, which shows the time-averaged ratio of gaseous species mass flow rates to the mass flow rates of those same species at equilibrium. The latter is computed by first evaluating the local equilibrium composition at every point in a given $x/d$ plane using constant enthalpy and pressure constraints in Cantera \cite{goodwin2020cantera}. The resulting equilibrium mass fractions are multiplied by the local gaseous mass flux and then summed over the full $x/d$ plane. The resulting ratio gives an indication of how close the bulk flow is to reaching chemical equilibrium.

\begin{figure*}[!h]
    \centering
    \includegraphics[width=\textwidth]{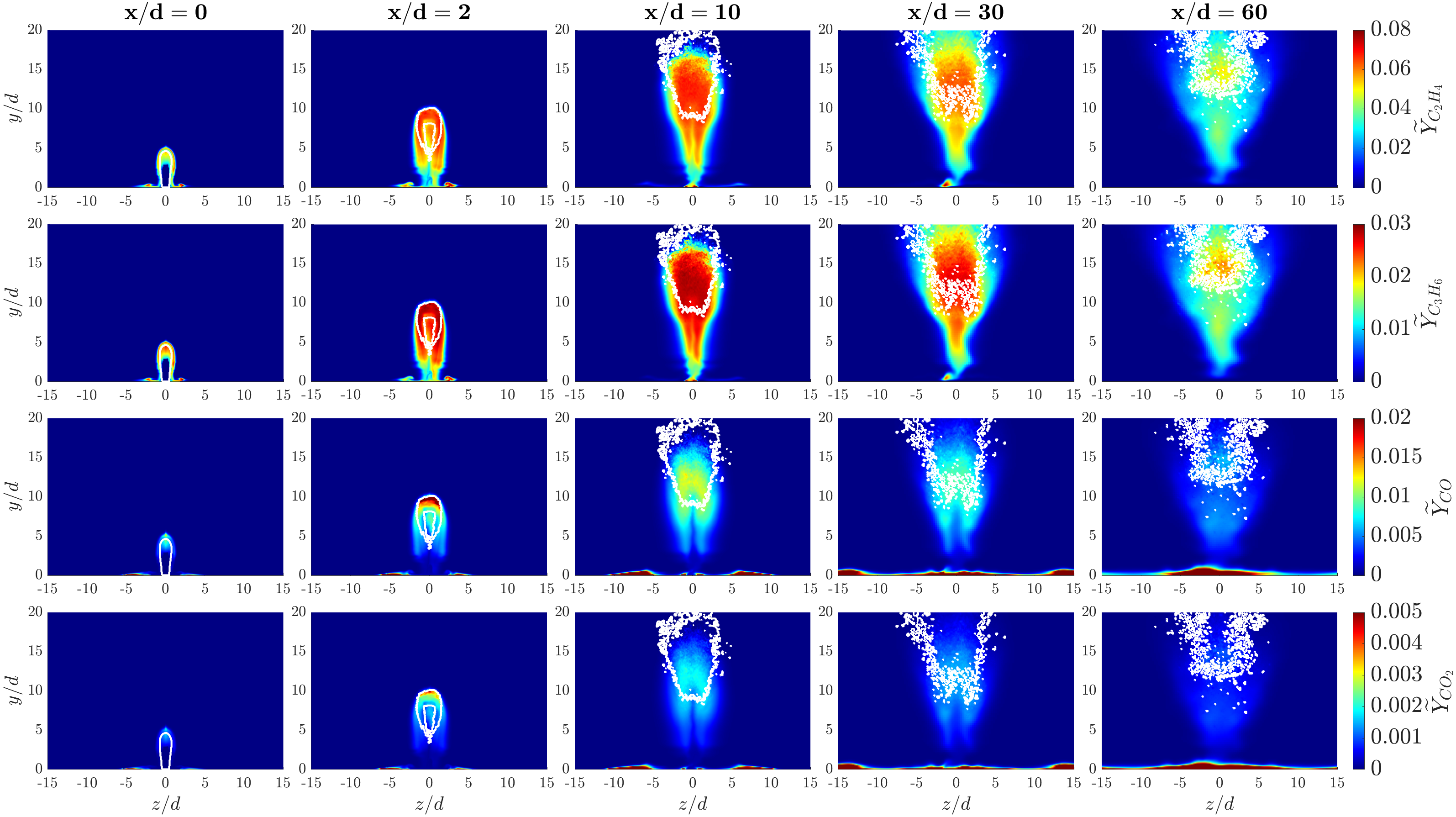}
    \caption{Favre-averaged gaseous species mass fractions at various $x/d$ planes in case \ref{case4_M8}. $\alpha_l>0.01$ isocontours colored in white.}
    \label{fig:Y_favre_x_case4_M8}
\end{figure*}

\begin{figure*}[!h]
    \centering
    \includegraphics[width=\textwidth]{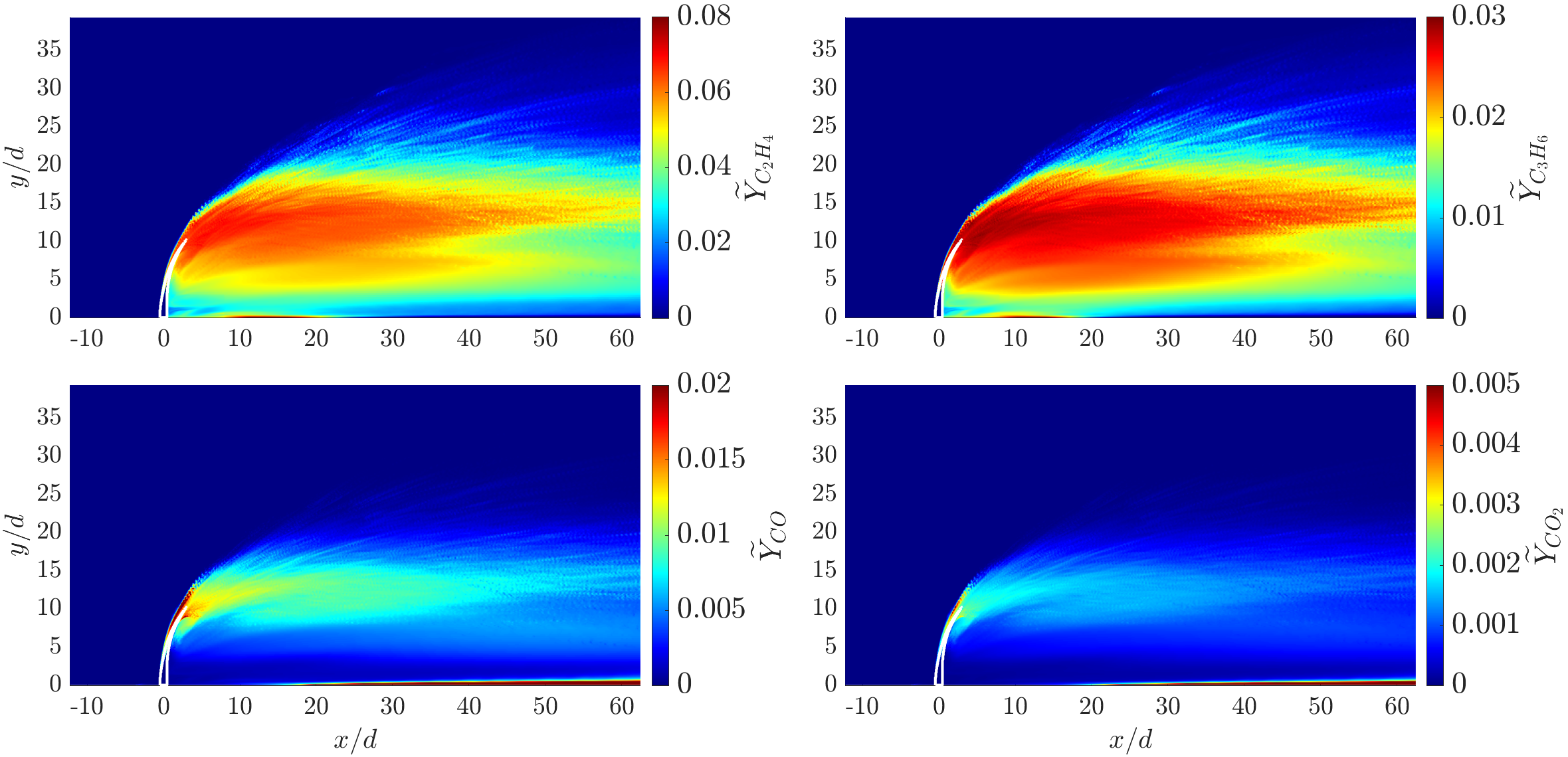}
    \caption{Favre-averaged gaseous species mass fractions along the $z$-midplane in case \ref{case4_M8}. $\alpha_l=1$ isocontours colored in white.}
    \label{fig:Y_favre_z_case4_M8}
\end{figure*}

As may be expected from Fig.\ \ref{fig:mdotx_Y__mdotx_g__timeavg}, the mass flow rates of intermediate species—namely, C$_2$H$_4$, C$_3$H$_6$, and C$_4$H$_8$—far exceed their equilibrium values for $x/d<20$ and settle closer to equilibrium farther downstream. Meanwhile, product species constitute a far lower proportion of the mass flux. Indeed, in case \ref{case2_M8} the ratios for the three dominant product species—CO, CO$_2$, and H$_2$O—are respectively 1.3\%, 1.2\%, and 6.0\% at $x/d=60$. In case \ref{case4_M8}, these ratios are respectively 0.6\%, 0.5\%, and 3.3\%. This illustrates that the bulk gaseous flow has not reached chemical equilibrium by the domain exit in either case. This can be largely attributed to the time required to reach equilibrium, which generally exceeds the residence time of the fuel. This can be seen in Fig.\ \ref{fig:eq_delay}, where the time to equilibrium for a stoichiometric n-dodecane-air mixture is plotted as a function of initial pressure and temperature. Here, the time to equilibrium is defined as the time required for the mass fractions of CO, CO$_2$, and H$_2$O to reach within 10\% of their equilibrium values in a zero-dimensional constant pressure reactor simulation in Cantera. Time is expressed in flow-through times, where one flow-through time is the time required for flow at the freestream velocity to traverse the distance from the injection point to the domain exit ($x/d \sim 60$). It is seen that the equilibrium composition is only obtained within one flow-through time for initial pressures and temperatures at the upper end of what is observed in the simulations. This region of $T-p$ space roughly corresponds to the region between the bow shock and liquid column in cases \ref{case2_M8} and \ref{case4_M8}. Granted, the lower speed regions in the jet wake can range between 20-40\% of the freestream velocity, so the timescales in Fig.\ \ref{fig:eq_delay} represent a worst-case scenario for the residence time. However, even when accounting for these lower velocities, the time required to reach equilibrium exceeds one of these lengthened flow-through times in much of the $T-p$ space. This analysis also does not account for the time required to atomize and evaporate the fuel, which will further delay the completion of the chemical reactions. As such, high-speed regions of the flow that dominate the mass fluxes in Figs.\ \ref{fig:mdotx_Y__mdotx_g__timeavg}-\ref{fig:mdotx_Y___mdotx_Y_eq__timeavg} are less likely to reach equilibrium by the domain exit.

This can be seen more directly in Figs.\ \ref{fig:Y_favre_x_case4_M8} and \ref{fig:Y_favre_z_case4_M8}, where Favre-averaged gaseous species mass fractions are respectively plotted on various $x/d$ planes and the $z$-midplane in case \ref{case4_M8}. The results for case \ref{case2_M8} showed similar features, so they have been omitted here for brevity. Both of these figures show that the intermediate species C$_2$H$_4$ and C$_3$H$_6$ reach their highest concentrations in the plume behind the liquid column ($5<y/d<20$). This is due to the high temperatures and pressures along the bow shock surrounding the windward side of the liquid column, which allow fuel to evaporate and begin to react. This can also be seen by the increased, albeit small, concentrations of CO and CO$_2$ along the front of the liquid column in Fig.\ \ref{fig:Y_favre_z_case4_M8}. However, farther downstream behind the liquid column, the temperatures remain comparatively low due to the evaporation of atomized fuel droplets (see Fig.\ \ref{fig:mach_T_HR_case4_M8}). As such, the fuel mass fraction continues to rise in this region, but there is minimal conversion to product species in the absence of substantial heat release. This leads to a dilution of the product gases in the jet wake, as can be seen along the $x/d=60$ plane in Fig.\ \ref{fig:Y_favre_x_case4_M8}.

However, as was shown in Figs.\ \ref{fig:mach_T_HR_case2_M8}-\ref{fig:mach_T_HR_case4_M8}, boundary layer heating facilitates higher temperatures in the low-velocity regions along the bottom wall. These factors respectively decrease the ignition delay and the distance required for the mixture to approach chemical equilibrium. As such, in Fig.\ \ref{fig:Y_favre_x_case4_M8}, intermediate species are observed within the counter-rotating vortices along the bottom wall at $x/d=0$ and $x/d=2$. Farther downstream, the reactions are not quenched by the evaporation of liquid fuel, as they are at greater heights within the jet wake. Correspondingly higher concentrations of final product species are thus observed farther downstream ($x/d>20$) as the mixture approaches equilibrium in the boundary layer ($y/d<2$). This helps to explain the behaviors illustrated in Figs.\ \ref{fig:mdotx_Y__mdotx_g__timeavg} and \ref{fig:mdotx_Y___mdotx_Y_eq__timeavg}. Due to the rapid reactions along the bow shock and subsequent quenching of reactions in the jet wake, the mass flow rates of several product species plateau after $x/d=20$. Farther downstream, only the boundary layer provides conditions conducive for chemical equilibration. Here, the mass flux is comparatively low, so the locally equilibrated mixture contributes little to the global combustion efficiency.

\begin{figure}[!h]
    \centering
    \includegraphics[width=\textwidth]{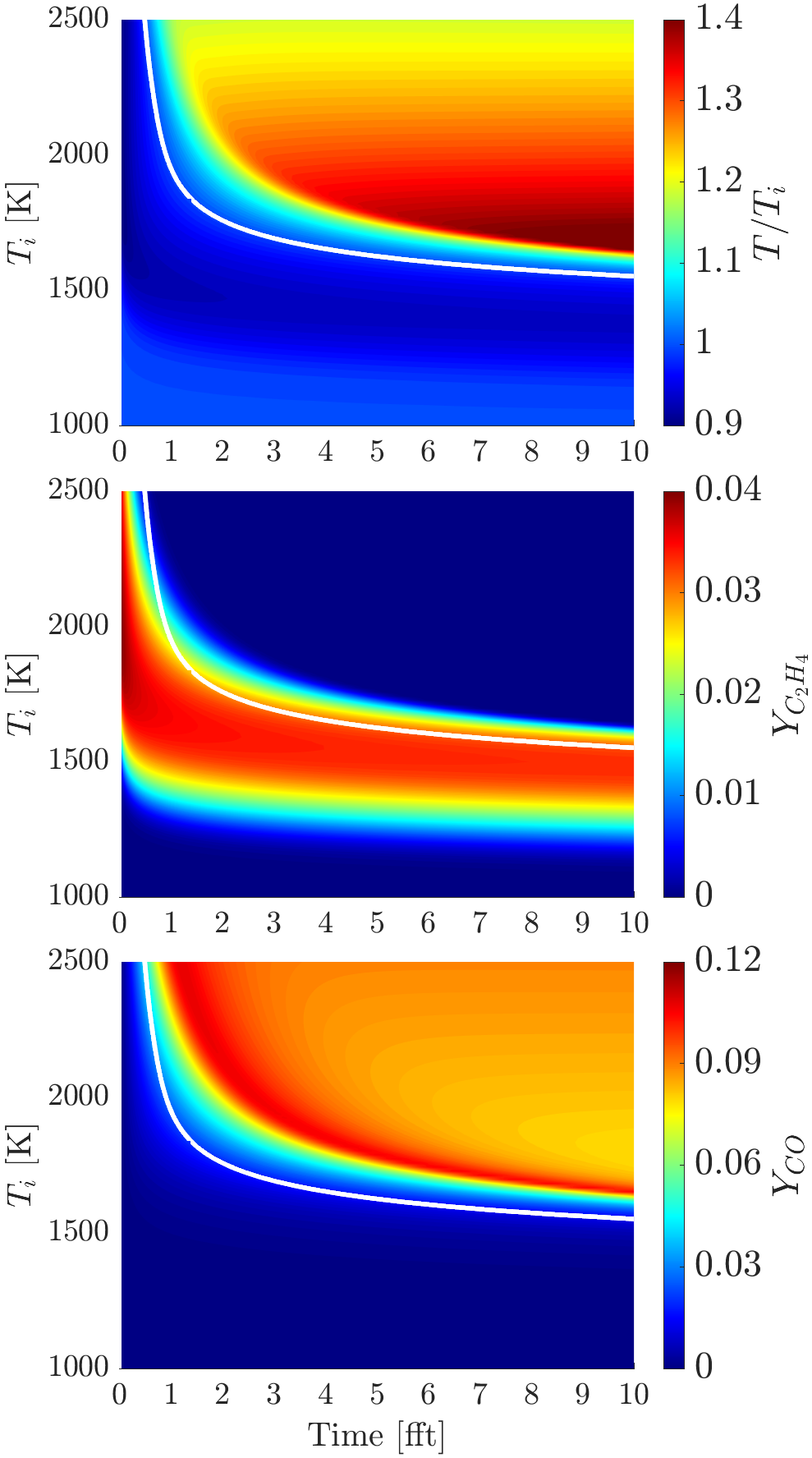}
    \caption{Temperature and product species in zero-dimensional constant-pressure reactor simulations with stoichiometric n-dodecane-air mixtures and $p_i=21.5$ kPa. White isoline marks $T/T_i=1$.}
    \label{fig:T_norm_Y_t_ft__p_fs_Y_st}
\end{figure}

\begin{figure}[!h]
    \centering
    \includegraphics[width=\textwidth]{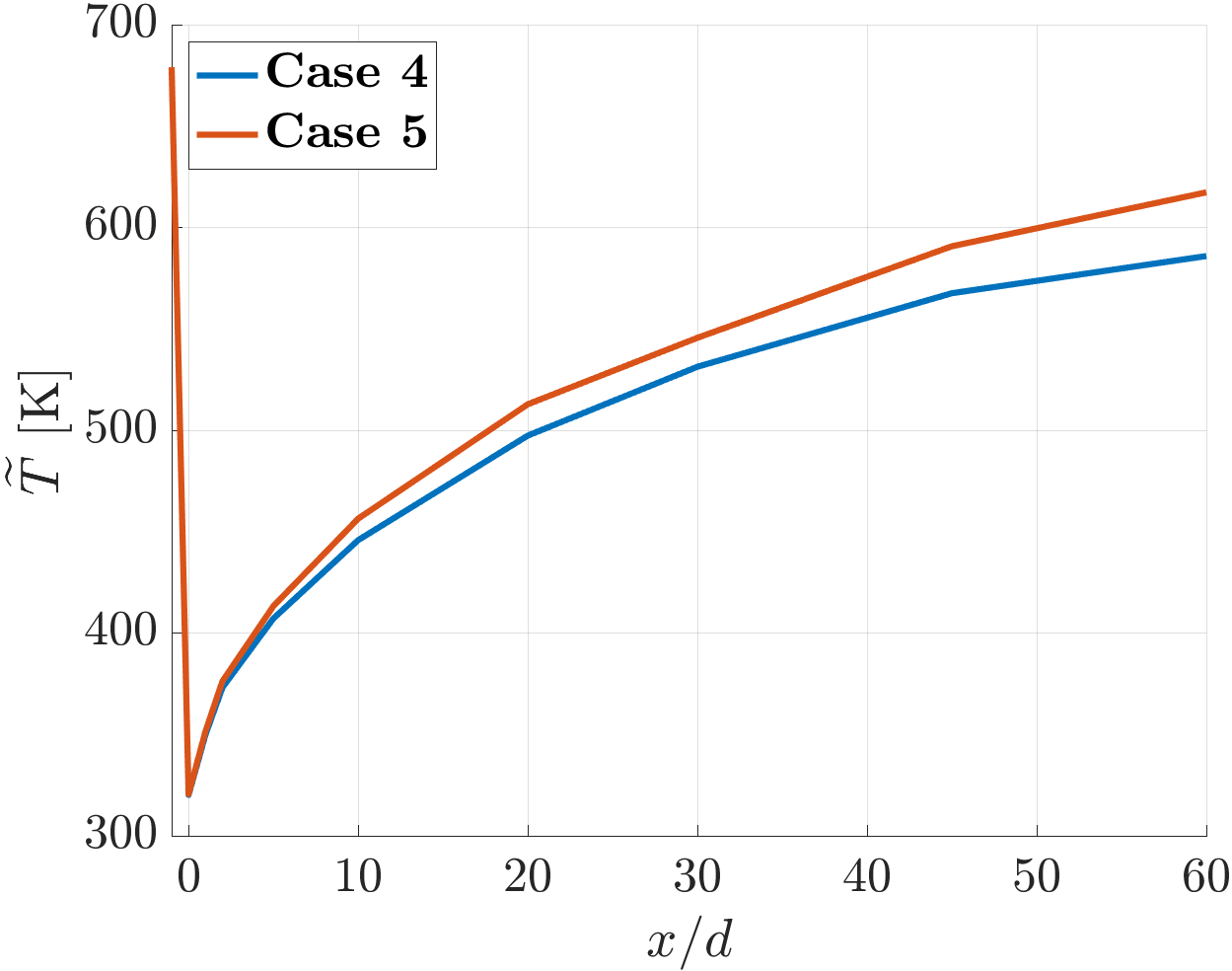}
    \caption{Favre-averaged temperatures across various $x/d$ planes in cases \ref{case4_M8} and \ref{case4_M8_nr}}
    \label{fig:T_favre_avg_xD_all}
\end{figure}

While not as significant as evaporation, it should be noted that endothermic combustion pathways can also act as a heat sink in these configurations. Figure \ref{fig:T_norm_Y_t_ft__p_fs_Y_st} shows the solutions through time of zero-dimensional constant-pressure simulations in Cantera using stoichiometric gaseous n-dodecane-air mixtures with varying initial temperatures $T_i$ and a constant initial pressure roughly equivalent to the freestream pressure in the current JISC simulations ($p_i = 21.5$ kPa). Here, time is again expressed in terms of flow-through times for the freestream flow. It can be seen that for low initial temperatures and short residence times, the final temperature can be on the order of 90\% of the initial temperature due to the endothermic cracking of n-dodecane molecules. This region of low temperature in these plots corresponds to high concentrations of intermediate product species, such as C$_2$H$_4$, similar to what is seen in the JISC simulations (Figs.\ \ref{fig:Y_favre_x_case4_M8}-\ref{fig:Y_favre_z_case4_M8}). Only with high initial temperatures—like those along the windward bow shock—or long residence times—like those in the leeward boundary layer—do final product species like CO appear in substantial quantities. This may contribute to a counterintuitive result obtained in cases \ref{case4_M8} and \ref{case4_M8_nr}, which is plotted in Fig.\ \ref{fig:T_favre_avg_xD_all}. Here, the Favre-averaged temperature is computed across the entirety of various $x/d$ planes for both cases.

The effect of the liquid's thermal mass is immediately apparent by the drop in $\widetilde{T}$ between $-1 < x/d < 0$ from the freestream temperature to nearly the liquid's injection temperature. Most notable, though, is that $\widetilde{T}$ is higher for the remainder of the domain length in the non-reacting case \ref{case4_M8_nr} than in the reacting case \ref{case4_M8}—peaking at a difference of 31 K at $x/d = 60$. It is unclear to what extent this is attributable to the endothermic fuel cracking shown in Fig.\ \ref{fig:T_norm_Y_t_ft__p_fs_Y_st}, as differences in boundary layer development and fuel jet atomization can also contribute to differences in temperature across the cross-section. However, this helps to illustrate the minimal effects of the chemical reactions in the present configurations and highlights a key feature when using more complex fuels in JISCs. Unlike other gaseous fuels commonly used in JISCs, such as H$_2$ and C$_2$H$_4$, the presence of endothermic fuel cracking produces additional cooling and potential combustion delays on top of those induced by evaporation of the liquid fuel.


\section{Conclusions \label{sec:conclusion}}

In this work, five simulations of reacting and non-reacting liquid jet in supersonic crossflow (JISC) configurations were conducted. Adaptive mesh refinement was utilized within a volume of fluid (VOF) methodology to capture the liquid jet atomization and turbulent mixing at high resolution. The freestream Mach number was held constant at roughly 4.5, while the liquid injection pressure and freestream temperature were varied to investigate the effects of the jet momentum ratio and freestream enthalpy on the jet penetration, mixing, and combustion dynamics.

The results show that similar jet penetration and mixing characteristics are observed for configurations with similar jet momentum ratios. The penetration heights are slightly lower than those in the experimental literature, but this was hypothesized to be due to a lack of boundary layer at the inflow. The greatest turbulent kinetic energy intensity is observed in the jet and atomization-induced shear layers, where vortex stretching and baroclinic torque are the primary generators of vorticity. Heat release is primarily observed along the leading bow shock and within the boundary layer in the jet wake. However, this heat release is not substantial enough to have an appreciable effect on the vorticity generation. Evaporative cooling in the primary atomization zone helps to quench the reactions at the injection height, such that the flow rates of several combustion products plateau after $x/d=20$. Substantial concentrations of final product species, such as CO, CO$_2$, and H$_2$O, are only observed along the leading bow shock and in the boundary layer far downstream of injection. The former is due to the locally elevated temperatures and pressures, which allow the flow to more rapidly reach chemical equilibrium, while the latter is due to the boundary layer heating and longer residence times in the near-wall flow. Overall, cooling from liquid fuel evaporation and, to a lesser extent, endothermic cracking of gaseous fuel, means that the short residence times afforded by the high-speed freestream are insufficient for the majority of the flow to reach chemical equilibrium prior to the domain exit. This manifests as low overall combustion efficiency for reacting cases with both lower and higher jet momentum ratios.

These results highlight the difficulties inherent in promoting substantial combustion in liquid JISC configurations. Future investigations can be conducted to develop means of facilitating faster fuel evaporation and shorter ignition delays, such as novel injection strategies or increasing the pressure and temperature at which the fuel and oxidizer interact. In any case, the additional multi-scale physical processes produced by liquid injection make these JISC setups more complex than their gaseous counterparts. Additional experimental and high-fidelity computational studies should be conducted to develop of more comprehensive understanding of these flows.


\section{Declaration of competing interest}

The authors declare that they have no known financial or personal relationships that could have appeared to influence the present work.

\section{Acknowledgments}

This work was supported by the US Office of Naval Research MURI N00014-22-1-2606, with Dr.\ Steven Martens as program manager. The computational resources were provided by the US Department of Defense High-Performance Computing Modernization Program (DoD HPCMP). The authors would like to acknowledge the efforts of Ral Bielawski and Lorenzo Angelilli in developing the multi-phase solver.


\bibliographystyle{elsarticle-num-names.bst}
\bibliography{references.bib}

\end{document}